\documentclass[letterpaper,twocolumn,10pt]{article}
\usepackage{usenix}

\usepackage{amsmath,amssymb,amsfonts}

\usepackage{graphicx}
\usepackage{textcomp}
\usepackage{xcolor}

\usepackage{xurl}
\usepackage{subfigure}
\usepackage{makecell}
\usepackage{algorithm}
\usepackage{algpseudocode}
\usepackage{tabularx}
\usepackage{diagbox}
\usepackage{enumitem}

\usepackage{xspace}
\usepackage{caption}
\usepackage{array}
\usepackage{booktabs}
\usepackage{array}
\usepackage[frozencache,cachedir=.]{minted} 
\usepackage{listings}
\usepackage{lipsum}

\newcommand\blfootnote[1]{%
  \begingroup
  \renewcommand\thefootnote{}\footnote{#1}%
  \addtocounter{footnote}{-1}%
  \endgroup
}

\newcommand{\re}[1]{\textcolor{black}{#1}\xspace}

\graphicspath{{Figs/}}

\algdef{SE}[DOWHILE]{Do}{doWhile}{\algorithmicdo}[1]{\algorithmicwhile\ #1}%
\def\NoNumber#1{{\def\alglinenumber##1{}\State #1}\addtocounter{ALG@line}{-1}}

\begin{document}

\newcommand{\sys}{PlanB\xspace}

\title
{
    \Large \bf \sys: Efficient Software IPv6 Lookup with Linearized $B^+$-Tree

}


\author{
{\rm Zhihao Zhang$^{1,3,4}$, Lanzheng Liu$^{1}$, Chen Chen$^{1}$, Huiba Li$^{1}$*, Jiwu Shu$^{4}$, Windsor Hsu$^{1}$, Yiming Zhang$^{2,3}$}\\
$^1$Alibaba Cloud \ \
$^2$NICE Lab, SJTU \ \
$^3$NICE Lab, XMU \ \
$^4$Tsinghua University
} 


\maketitle

\begin{abstract}

IP lookup via Longest Prefix Match (LPM) is critical for packet forwarding.
Unfortunately,
conventional lookup algorithms
are inefficient for IPv6 Forwarding Information Bases (FIBs),
which are characterized by a set of long prefixes with diverse lengths.
We observe that LPM inherently represents a two-dimensional (2D) search problem
over both prefix values and prefix lengths,
but
existing algorithms mostly
treat LPM as two separate levels of one-dimensional (1D) searches,
causing poor lookup performance and high memory overhead.

This paper presents PlanB,
a novel scheme for high-speed IPv6 lookup. 
We transform the 2D LPM into an equivalent 1D search problem over elementary intervals,
thereby 
unifying the search across prefix value and lengths.
We then adapt a flat-array-based B-tree structure to the needs of LPM 
to propose the \emph{linearized $B^+$-tree},
based on which 
we introduce an efficient search algorithm 
tailored to the properties of the transformed space.
To maximize performance,
we integrate \sys with vectorization, batching, branch-free logic, and loop unrolling to fully exploit CPU parallelism.
Extensive evaluation 
shows that
\sys achieves single-core performance of 390 Million Lookups Per Sec (MLPS) with real-world IPv6 FIBs on AMD processor,
and scales to full-12-core performance of 3.4 Billion Lookups Per Sec (BLPS).
This is 1.6$\times$$\sim$14$\times$ higher than state-of-the-art software-based schemes (PopTrie, \re{CP-Trie, Neurotrie} and HBS).

\blfootnote{*Huiba Li (huiba.lhb@alibaba-inc.com) is the corresponding author.}




\end{abstract}



\vspace{-3mm}
\section{Introduction}
\label{sec:intro}

The Internet's transition to IPv6 is accelerating, 
driven by 
the exhaustion of IPv4 addresses and the increasing connectivity demands of IoT (Internet of Things) and mobile networking. 
The transition fuels unprecedented growth in global IPv6 traffic and routing table sizes,
with the latter projected to exceed 300,000 prefixes by 2030 \cite{RouteViews,RIPE-RIS,IPv6-BGP-Data}.
This trend is evident across the globe.
Google reports that over 45\% of its users now access services via IPv6 \cite{Google-IPv6-Statistics}, and adoption rates in countries like the United States, Germany, and India have exceeded 57\% \cite{APNIC-IPv6-Statistics}. 
The transition is particularly pronounced in countries facing IPv4 address shortages.
For instance,
China now has over 822 million IPv6 users \cite{CNNIC-Report},
with IPv6 constituting over 31\% of its national network traffic \cite{IPv6-Development-China}.
Moreover,
it coincides with a broader shift in networking towards
Network Function Virtualization (NFV) \cite{Kablan-HotMiddlebox,NFP-SIGCOMM,Bento-SIGCOMM,GreenNFV-SC,ResQ-NSDI}.
The reliance of these architectures on software packet processing makes the efficiency of IP lookups especially critical.
The rapid adoption of IPv6,
coupled with longer prefixes, larger routing tables,
and increasing traffic volumes,
places significant pressure on existing lookup mechanisms.


Hardware-based IP lookup solutions, leveraging components like TCAMs \cite{NeuroLPM-MICRO, TCache-ToN, Sadeh-SOSR,CATCAM-MICRO,casado2008rethinking}, FPGAs \cite{Stimpfling-CCGRID,SHIP-ToN,DHL-ICDCS}, and ASICs \cite{CP-Trie,C2RTL-HPSR}, offer high throughput but face significant limitations.
These specialized components are prohibitively expensive and consume significant power,
and their limited capacity is insufficient for today's rapidly growing IPv6 routing tables \cite{NeuroLPM-MICRO,CATCAM-MICRO,Jiang-INFOCOM,Zheng-TON},
making them impractical for large-scale or cost-sensitive deployments \cite{Mogul-PhysicalDeployability,Bhowmik-DEBS,Bando-ToN}.
Furthermore, specialized hardware lacks flexibility required by modern paradigms like NFV \cite{Orion-NSDI,Bento-SIGCOMM,ResQ-NSDI},
which depend on software for rapid service deployment and management.
As a result, there is a trend towards implementing routers in software on commodity servers \cite{SWAN,Shao-NSDI,RouteBricks-SOSP}.
Software-based IPv6 lookup solutions are more cost-effective and flexible for cloud deployments.

However, software-based solutions face a long-standing challenge: IP lookup remains a critical performance bottleneck. The problem is particularly severe for IPv6, where the characteristics of Forwarding Information Bases (FIBs) significantly degrade router performance. 
Specifically,
IPv6 FIBs contain a wide range of prefix lengths \cite{Piraux-SIGCOMM,Rye-SIGCOMM,Luori-ICNP,AddrMiner-ATC,TAR-APNet}, with typical backbone entries spanning from /32 to /64,
increasing both computational overhead and memory footprint.
Compared to IPv4 \cite{HBS-ToN,Li-ICICSP},
this results in a 3$\times$--10$\times$ higher lookup cost for state-of-the-art algorithms
and a 10$\times$--50$\times$ increase in memory footprint.

Existing software-based IP lookup schemes,
which struggle to meet the dual demands of high speed and low memory footprint for IPv6 \cite{Neurotrie-ToN,CP-Trie,Li-ICICSP}, fall into two main categories.
First,
trie-based schemes, 
such as SAIL\cite{SAIL-SIGCOMM,SAIL-ToN} and Poptrie \cite{Poptrie},
construct a multi-bit trie to traverse the prefix space.
These methods use fixed strides (e.g., 6 bits per level) or fixed partitioning schemes to reduce memory accesses.
Second,
hash-based schemes,
such as HBS \cite{HBS-ToN} and HHR \cite{Li-ICICSP},
perform a binary search (BS) over the set of unique prefix lengths,
using hash tables at each step to check for a prefix match.

IP lookup determines the forwarding path by selecting the longest prefix match (LPM) for a given destination address.
The performance of existing lookup schemes is limited by a key problem: they treat LPM, an inherently two-dimensional (2D) search problem across prefix values and lengths, as two separate one-dimensional (1D) searches.
This separation is inefficient for large IPv6 FIBs.
The problem is further compounded by memory hierarchy constraints.
To achieve high throughput,
the lookup structure must fit into the fast but capacity-limited SRAM (i.e., CPU caches) \cite{SAIL-ToN,Poptrie,TAR-APNet},
rather than being stored in the larger but slower DRAM.
As a result, efficient IPv6 lookup schemes need to be 
cache-friendly.

In this paper, we present \sys, a novel IPv6 lookup scheme designed to achieve high lookup performance on commodity hardware.
\re{\sys is distinguished by its integration of a 2D-to-1D dimensional reduction with a pointer-less, cache-aligned linearized $B^+$-tree layout.
This combination specifically enables the data-parallel vectorization that standard trie-based structures cannot support.}
\sys addresses the inefficiencies of existing methods in three aspects.

First,
\sys transforms the 2D search problem of LPM into a 1D search problem.
\sys reframes this by representing each prefix/prefix\_length entry as a range [start\_address, end\_address].
\sys partitions the entire address space into a set of non-overlapping, elementary intervals.
This transformation elegantly converts the complex 2D problem into a simple 1D search on the sorted start addresses of these intervals.

Second,
to efficiently solve the resulting 1D search problem,
\sys proposes linearized $B^+$-tree,
which adapts the flat-array B-tree structure \cite{Khuong-ACM-JEA,Sherman,eurosys25deft,uTree} to the specific needs of LPM
and maps the entire $B^+$-tree into a single, contiguous array
thus being inherently pointer-less.
Based on linearized $B^+$-tree,
\sys's search algorithm navigates the tree by computing parent-child relationships using simple arithmetic on array indices,
and employs a lower-bound binary search \footnote[1]{The lower-bound binary search finds the first element in a sorted array that is not less than the given target value.}
at each node to find the appropriate child to traverse next.
The linearized $B^+$-tree eliminates pointer chasing,
increases data density,
and aligns nodes with cache lines.



\begin{table*}[t]
    \centering
    \small
    \caption{\label{table:datasets}
        The details of collected FIBs from RIPE and RouteViews.
    }
    \vspace{-2mm}
    \renewcommand{\arraystretch}{1.2}
        \begin{tabularx}{0.94\textwidth}{p{1.1cm}>{\centering\arraybackslash}p{1.2cm}>{\centering\arraybackslash}p{2.7cm}>{\centering\arraybackslash}X>{\centering\arraybackslash}p{1.9cm}}
            \hline
            \textbf{Name} & \textbf{Time} & \textbf{Physical Location} & \textbf{IP Prefix Distribution (Top 5)} & \textbf{\# of Prefixes} \\
            \hline
            rrc00-19     & 20190801   & Amsterdam, NL   & 48(46.52\%),32(17.13\%),44(6.13\%),40(5.40\%),36(3.97\%)          & 76342           \\ \hline
            rrc00-20     & 20200801   & /               & 48(47.87\%),32(15.52\%),44(6.14\%),40(5.31\%),36(4.15\%)          & 95504            \\ \hline
            rrc00-21     & 20210801   & /               & 48(41.82\%),32(14.59\%),44(8.73\%),40(6.66\%),64(4.81\%)           & 143823              \\ \hline
            rrc00-22     & 20220801   &  /              & 48(47.28\%),32(13.31\%),44(8.32\%),40(7.73\%),36(3.76\%)         & 168572                  \\ \hline
            rrc00-23     & 20230801   &  /              & 48(46.39\%),32(11.83\%),44(9.09\%),40(7.89\%),36(3.65\%)         & 199801                  \\ \hline
            rrc00-24     & 20240801   & /               & 48(45.27\%),32(11.19\%),44(9.40\%),40(9.01\%),36(3.64\%)         & 226222                   \\ \hline
            rrc00-25     & 20250801   &  /              & 48(44.55\%),32(11.00\%),40(10.11\%),44(9.69\%),36(3.89\%)         & 235466
            \\ \hline
            rv1          & 20250801   & Johannesburg, SA & 48(44.78\%),32(11.07\%),40(10.00\%),44(9.47\%),36(3.76\%)        & 235038                  \\ \hline
            rv2          & 20250801   & Singapore, SG  & 48(44.93\%),32(10.92\%),40(9.85\%),44(9.56\%),36(3.71\%)     & 238498                   \\ \hline
            rv3          & 20250801   & Tokyo, JP   & 48(46.03\%),32(11.42\%),40(10.30\%),44(9.19\%),36(3.83\%) & 226893                    \\ \hline
            rv4          & 20250801   & New York, USA   & 48(44.88\%),32(11.30\%),40(10.17\%),44(9.69\%),36(3.82\%) & 229417                 \\ \hline
            rv5          & 20250801   & Rio de Janeiro, BR  & 48(45.37\%),32(11.26\%),40(9.36\%),44(9.24\%),36(3.86\%) & 227833                   \\ \hline
            rv6          & 20250801   & Sydney, AU   & 48(44.92\%),32(11.21\%),40(10.15\%),44(9.53\%),36(3.80\%) & 230672                   \\ \hline
            rv7          & 20250801   & Frankfurt, DE   & 48(44.56\%),32(11.02\%),40(10.02\%),44(9.49\%),36(3.82\%) & 234916                   \\ \hline  
            \end{tabularx}
\vspace{-3mm}
\end{table*}

Third, \sys implements the search algorithm with a set of optimizations.
It leverages Single Instruction Multiple Data (SIMD) instructions (e.g., AVX-512)
and batching to enable simultaneous comparisons of the target address against multiple keys within a $B^+$-tree node.
The search is further designed to be branch-free:
by replacing conditional statements with bitwise operations and masks,
\sys eliminates costly branch misprediction penalties and maintains a more efficient CPU pipeline.
In addition, compile-time loop unrolling reduces loop-control overhead, further accelerating the search.





By combining these techniques,
\sys delivers high-speed IPv6 lookups on commodity hardware.
Our evaluation using real-world IPv6 FIBs on a 12-core AMD processor
shows that 
\sys achieves 390 Million Lookups Per Sec (MLPS) on a single core. 
The throughput scales linearly with the core count, reaching 3.4 Billion Lookups Per Sec (BLPS) across all cores,
significantly outperforming existing state-of-the-art software solutions.
We further test \sys with synthetic FIBs containing up to one million prefixes,
showing that \sys always achieves high performance.
\re{Crucially, \sys makes no distribution assumptions, and its effectiveness on both real-world and synthetic datasets confirms its robustness.
Furthermore,
\sys demonstrates near constant performance under fully random traffic. This resilience stems from a robust adaptability to dynamic workloads that inherently lack exploitable temporal locality. Because search latency remains strictly bounded and independent of traffic predictability, the system sustains optimal throughput even during severe traffic bursts.
For various IPv6 FIBs,
\sys is 1.6$\sim$14 times faster than state-of-the-art software-based schemes, PopTrie, CP-Trie, Neurotrie and HBS. Additionally, it reduces memory overhead by 56.4\%$\sim$92.5\%.}

\vspace{-3mm}
\section{Background and Motivation}
\label{sec:background}







\subsection{IPv6 Prefix Length and Distribution}
\label{sec:ipv6-prefixes}

For every packet it forwards, a router performs an IP lookup to determine the correct next hop. This operation queries a FIB, which contains a set of (prefix, next\_hop) rules. Given a packet's destination address, the router must find the rule with the longest matching prefix \cite{NeuroLPM-MICRO,SAIL-ToN}.
This matching rule dictates where the packet should be sent next. While the LPM principle is the same for both IPv4 and IPv6, the shift to IPv6 introduces significant challenges that fundamentally alter the performance landscape of lookup algorithms.

To understand the challenges of IPv6 routing,
we collect seven large real-world backbone network datasets from RIPE \cite{RIPE-RIS} and seven backbone datasets from RouteViews \cite{RouteViews} to analyze prefix length and distributions.
The RIPE datasets are obtained from the RRC00 collector at 8:00 AM,
covering the first day of August for each year from 2019 to 2025.
The RouteViews datasets are collected from multiple geographic regions, including Africa, Asia, North and South America, Australia, and Europe.
All datasets are processed using the Zebra FIB converter \footnote[2]{https://github.com/rfc1036/zebra-dump-parser} and bgpdump \footnote[3]{https://github.com/RIPE-NCC/bgpdump} to remove duplicate rules,
ensuring they can be directly utilized by our evaluation framework.
A detailed summary of these datasets is provided in Table \ref{table:datasets}.
Our analysis of these datasets reveals several key characteristics of IPv6 FIBs that motivate the need for a new lookup algorithm.

\textbf{Rapid FIB Growth and Increasing Granularity.}
The IPv6 routing table is not only growing but is doing so at a dramatic pace as shown in Table \ref{table:datasets}.
The number of prefixes in the RIPE dataset increases from 76,342 in 2019 to a projected 236,466 in 2025—a more than threefold increase in just six years.
This rapid scaling demands a lookup algorithm with low memory overhead and high throughput.
More importantly,
the structure of the FIB is evolving.
While the proportion of /48 prefixes remains high,
we observe a significant trend towards de-aggregation.
From 2019 to 2025, the share of aggregated /32 prefixes drops from 17.13\% to 11.00\%.
Concurrently,
the shares of more specific, longer prefixes rise;
for instance, /40 prefixes nearly double their representation from 5.4\% to 10.11\%.
This shift towards finer granularity deepens the lookup problem, as algorithms must efficiently handle a growing number of longer prefixes.

\textbf{Highly Skewed and Concentrated Distribution.}
A defining feature of IPv6 FIBs is their highly skewed prefix-length distribution.
IPv6 routing tables are dominated by a few specific prefix lengths.
Across all 14 datasets, /48 is the most common prefix length, consistently accounting for 41-48\% of all entries.
The prevalence of these long prefixes makes traditional bit-wise trie traversal inefficient, as it requires navigating deep data structures.
The 2025 RouteViews data shows that this skewed distribution is remarkably consistent across the globe.
From Johannesburg to Tokyo,
the total prefix count varies by less than 6\%,
and the proportions of dominant prefix lengths are nearly identical.
For example, the share of /48 prefixes only varies between 44.5\% and 46.0\%.
Moreover, Our additional observation is the concentration of prefixes. Across all datasets, just five prefix lengths, /32, /36, /40, /44, and /48, consistently comprise approximately 80\% of the entire FIB. For example, in the rrc00-25 and rv1 datasets, these five lengths account for 79.2\% and 79.1\% of prefixes, respectively.

The combination of fast table growth, skewed prefix lengths, and global uniformity presents key requirements: (i) Algorithms should be designed around the reality that nearly half of lookups are against /48s, (ii) Lookup performance should be insensitive to the table expanding by hundreds of thousands of entries within a few years, and (iii) Since distributions are stable both temporally and geographically, an adaptive algorithm tuned to the observed mix of longer prefixes will remain effective globally and into the near future.

\vspace{-3mm}
\subsection{Software-based Lookup Methods}
\label{sec:software-lookup-methods}

Software-based IP lookup schemes can be mainly divided into trie-based \cite{Poptrie, SAIL-SIGCOMM,SAIL-ToN} and hash-based \cite{Hi-BST,HBS-APNet,HBS-ToN} methods.

\textbf{Trie-based methods} construct a tree where the path from the root represents an IP prefix.
While effective for IPv4,
their performance on IPv6 is often hampered by the increased address length.
SAIL \cite{SAIL-SIGCOMM,SAIL-ToN} separates the lookup process into finding the prefix length and the next-hop, and it splits prefixes by length (e.g., $\leq24$ vs. $>24$).
By expanding prefixes and using large bitmap arrays,
SAIL achieves a constant, low number of memory accesses for most IPv4 traffic.
Its primary drawback is a massive memory footprint.
For IPv6 which prefix lengths are typically less than 64 bits in backbone rounters,
the tatol memory size of all bitmaps in SAIL is $\sum_{i=0}^{64} 2^i = 4EB$.
While it works for the shorter prefixes of IPv4,
its memory requirement explodes exponentially with prefix length,
rendering it unusable for the larger address space of IPv6.
The memory footprint of SAIL exceeds the typical CPU cache size,
requiring relatively slow DRAM access in case of cache misses.
Moreover,
the performance of SAIL relies on the destination IP address locality of the traffic pattern

PopTrie \cite{Poptrie} is a state-of-the-art trie-based algorithm that builds a compressed 64-ary trie.
It achieves a small memory footprint and high lookup speed for IPv4 by using bitmaps and the popcnt CPU instruction to quickly identify child nodes.
However, its fixed 6-bit stride is a critical limitation for IPv6.
A 48-bit prefix requires traversing up to 8 levels in the worst case, leading to numerous memory accesses that diminish cache efficiency and degrade performance.
CP-Trie \cite{CP-Trie} extends PopTrie by incorporating a cumulative popcount, which allows for longer fixed strides while still using a single popcnt instruction per step.
This reduces the trie depth compared to PopTrie,
resulting in fewer memory accesses.
Nonetheless,
because its stride is still fixed,
it cannot adapt its structure to the specific prefix distribution of a given FIB,
making it suboptimal for the diverse IPv6 landscape.
\re{Neurotrie \cite{Neurotrie-ToN} introduces an adaptive trie structure that supports arbitrary strides at each node.}
It uses deep reinforcement learning (DRL) to determine the optimal stride for each node,
aiming to minimize trie depth while respecting a memory budget.
Its main weakness is the construction process,
which is computationally intensive and time-consuming,
requiring significant training.

\textbf{Hash-based methods} partition prefixes by length into separate hash tables. The classic Binary Search on Prefix Lengths (BS) scheme and its derivatives \cite{Hi-BST,HBS-APNet,HBS-ToN} organize these hash tables into a Binary Search Tree (BST) to guide the lookup.
HBS (Heuristic Binary Search) \cite{HBS-ToN} is a modern BS-family scheme designed for IPv6.
It enhances the classic BS approach with several key techniques.
First, it employs a heuristic search that uses pre-computed information about marker origins to prune the search space within the BST,
reducing the number of hash probes.
Second,
it uses tree rotation to dynamically adjust the BST structure in response to changes in prefix distribution,
ensuring the tree remains optimized without full reconstruction.
However,
its worst-case lookup time is not as tightly bounded as in trie-based schemes, and its performance depends on the hash function and the dynamic shape of the BST.


\begin{figure}[t]
  \centering
  \includegraphics[width=0.99\linewidth]{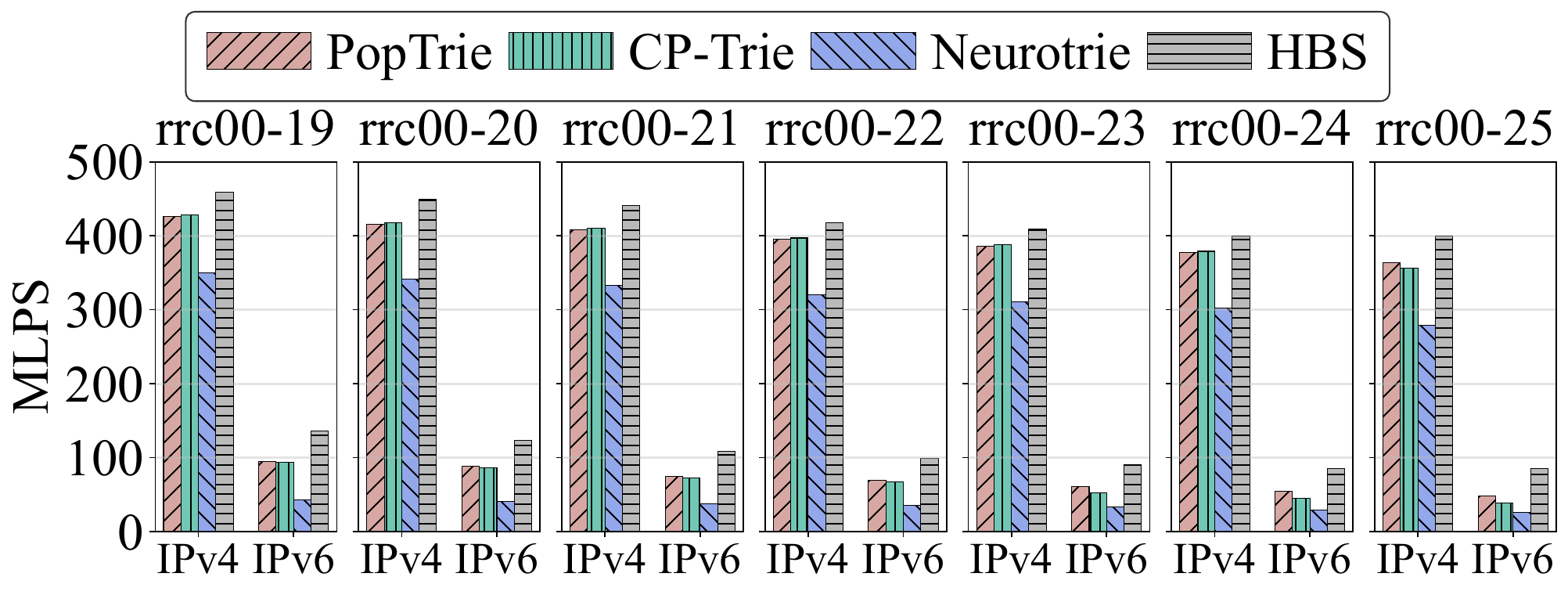}
  \vspace{-3mm}
  \caption{Lookup performance of four software schemes.}
  \label{fig:mot_mlps}
  \vspace{-3mm}
\end{figure}

We implement and benchmark several state-of-the-art software-based IP lookup methods.
We then evaluate these methods on the Intel server described in Section \ref{sec:evaluation}.
We use the real-world IPv6 FIBs detailed in Table \ref{table:datasets} for our benchmarks.
\re{Fig. \ref{fig:mot_mlps} presents the lookup performance in Million Lookups Per Second (MLPS) of PopTrie, CP-Trie, Neurotrie and HBS.}
A stark and immediate observation is the dramatic performance degradation from IPv4 to IPv6 across all tested methods.
For instance, in the rrc00-19 dataset,
the lookup speed for the schemes drops by over 75\% when transitioning from IPv4 to IPv6.
For both schemes, the lookup cost for IPv6 is significantly higher (3$\times$ to 10$\times$) than for IPv4.
This highlights the fundamental challenge posed by the IPv6's address space,
which increases structure size and processing overhead,
thereby reducing cache efficiency and overall throughput.

Furthermore,
the figure reveals a consistent decline in performance for all algorithms as the FIB size increases from 2019 to 2025.
For example,
the IPv6 lookup speed of HBS drops by 37.5\%, from 136 MLPS on the rrc00-19 dataset to 85 MLPS on the rrc00-25 dataset.
This trend underscores the scalability challenge that all software-based lookup schemes face as routing tables continue to grow.
Moreover,
The memory hierarchy is a critical factor in determining lookup performance.
Once the FIB size exceeds the CPU's combined cache capacity (36 MB on our test platform),
all algorithms suffer noticeable slowdowns due to the increased cost of off-chip DRAM accesses.
Thus,
an effective lookup data structure must not only fit within fast but limited on-chip SRAM, but also be highly cache-efficient.

\vspace{-3mm}
\section{Design and Implementation}
\label{sec:design}


As established in Section \ref{sec:background},
existing software-based IPv6 lookup schemes fail to scale with the size and complexity of modern FIBs.
To overcome these limitations,
we propose \sys,
a new framework that reformulates the LPM as a one-dimensional search problem over elementary intervals, resolved efficiently using a linearized $B^+$-tree. 

In this section, we first outline the design principles of \sys,
then describe its core components and optimizations in detail.
We assume \sys is used solely to obtain the FIB index for next-hop selection during IP forwarding.
The actual routes are maintained separately in a routing table (RIB),
such as a radix \cite{ART-ICDE} or Patricia trie \cite{Poptrie},
allowing aggressive compression of routes that share the same next hop.

\vspace{-2mm}
\subsection{From 2D Prefixes to 1D Intervals}
\label{sec:one-dimensional-search}

\sys transforms the LPM into a one-dimensional search problem.
This transformation is performed as a pre-computation process each time the FIB is updated.
The process involves two key steps.

\begin{figure}[t]
  \centering
  \includegraphics[width=0.91\linewidth]{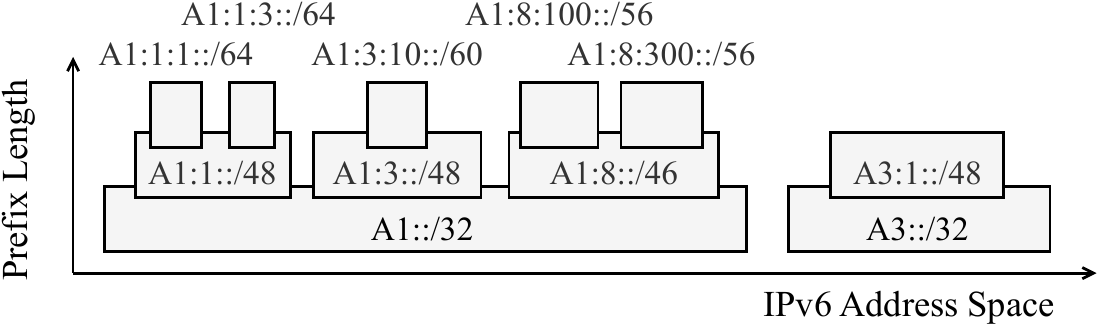}
  \vspace{-2mm}
  \caption{Visualization of IPv6 prefixes as prioritized address ranges. The horizontal and vertical axes represent the IPv6 address space and prefix length (priority), respectively.
  }
  \label{fig:design-1}
  \vspace{-3mm}
\end{figure}

\textbf{Range Conversion.}
Each forwarding rule,
defined by a prefix $P$ and a prefix length $L$,
is converted into a closed interval $\left[start, end\right]$ on the 128-bit unsigned integer space.
The $start$ address is simply the prefix $P$ itself,
while the $end$ address is calculated as $P + (2^{128-L} - 1)$.
For example,
the prefix 2001:0db8::/32 is converted to the range [2001:0db8::, 2001:0db8:ffff:ffff:ffff:ffff:ffff:ffff].
This conversion maps all prefixes, regardless of their length,
into a uniform format of address ranges. 
As visualized in Fig. \ref{fig:design-1},
this can be conceptualized in two dimensions:
the x-axis represents the entire IPv6 address space,
and the y-axis represents priority.
Each rule becomes a rectangle whose width corresponds to its address range and whose height corresponds to its prefix length (priority).
Consequently,
more specific routes (longer prefixes) are represented as narrower,
higher-priority rectangles that visually overlay the wider,
lower-priority rectangles of less specific routes.

\begin{figure}[t]
  \centering
  \includegraphics[width=0.91\linewidth]{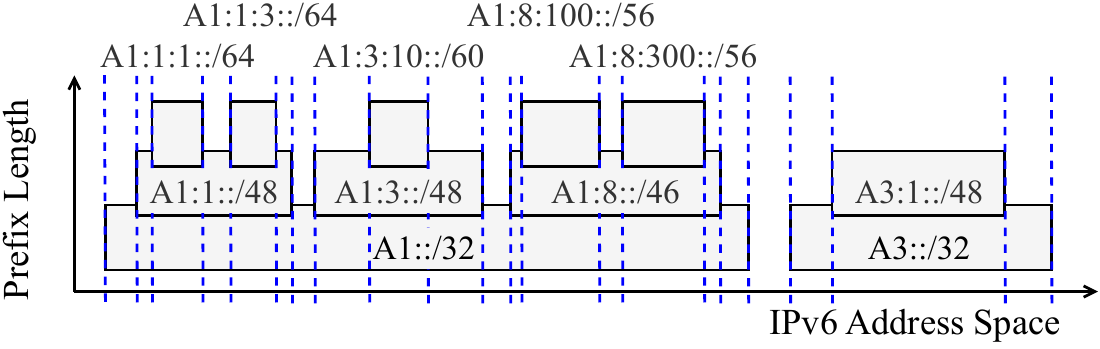}
  \vspace{-2mm}
  \caption{Partitioning the address space into non-overlapping elementary intervals using the start and end points of prefix ranges as boundaries (vertical dashed lines).
  }
  \label{fig:design-2}
  \vspace{-2mm}
\end{figure}

\textbf{Interval Partitioning.}
Next,
the $start$ and $end$ addresses from all converted ranges are collected.
These boundary points are sorted and used to partition the 128-bit IPv6 address space into a set of disjoint, contiguous elementary intervals.
\re{This partitioning process is deterministic and requires no tunable parameters, as the partitions are derived directly from the prefix boundaries.}
As illustrated by the vertical dashed lines in Fig. \ref{fig:design-2},
these boundaries divide the address space into non-overlapping segments.
For a FIB containing $N$ prefixes,
this process generates $2N$ boundary points ($N$ start points and $N$ end points), 
resulting in at most $2N + 1$ elementary intervals.

This procedure constructs a canonical,
one-dimensional data structure from the FIB,
implemented as a sorted array of elementary intervals.
Each element in the array associates a best-matching route (next-hop) pre-calculated along with the partitioning.
The elements that are not covered by any prefix are associated with the default route.
Consequently,
a lookup for a destination address $D$ is reduced to a standard predecessor search: 
finding the interval in the sorted list with the largest $interval\_start$ value that is less than or equal to $D$.
This reduction of two-dimensional LPM to a one-dimensional search problem is fundamental to \sys's efficiency.
The 1D transformation is distribution-agnostic as it operates on the boundaries of prefix ranges, not the prefixes themselves.
The critical insight of this partitioning is that all addresses within a single elementary interval are covered by the same set of original prefix ranges.
Due to the LPM rule, a single, highest-priority route (i.e., the one with the longest prefix length) is the definitive best match for all addresses in that interval.


\vspace{-2mm}
\subsection{1D Search with Linearized $B^+$-Tree}


To efficiently support high-speed lookups, \sys proposes linearized $B^+$-tree, which extends the flat-array layout originally developed for standard B-tree structures \cite{Khuong-ACM-JEA}.

\begin{figure}[t]
  \centering
  \includegraphics[width=0.99\linewidth]{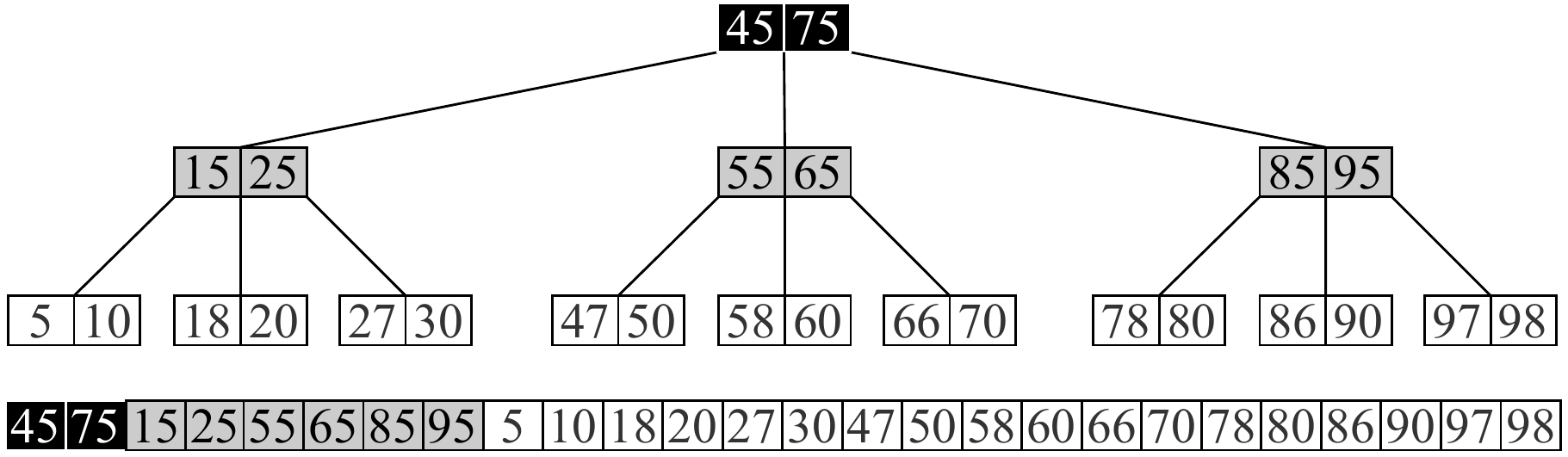}
  \vspace{-1mm}
  \caption{An example of the flat array layout of a 3-ary B-tree.
  }
  \label{fig:linear-b-tree}
  \vspace{-3mm}
\end{figure}

\textbf{Flat Array Layout of B-Tree.}
A B-tree is a balanced tree structure in which each node contains multiple keys and children \cite{Comer-ACM-CS}.
In the flat array layout of B-tree \cite{Khuong-ACM-JEA},
the data are conceptually organized as a complete binary search tree.
The values of the nodes in this virtual tree are stored in an array in the order they appear during a left-to-right breadth-first traversal (Fig. \ref{fig:linear-b-tree}).
However,
a key limitation of this approach is that the search process must continuously check if the target key has been found at each internal node.
These intermediate checks introduce conditional branches,
which incur lookup overhead and complicate acceleration through batching.

\begin{figure}[t]
  \centering
  \includegraphics[width=0.99\linewidth]{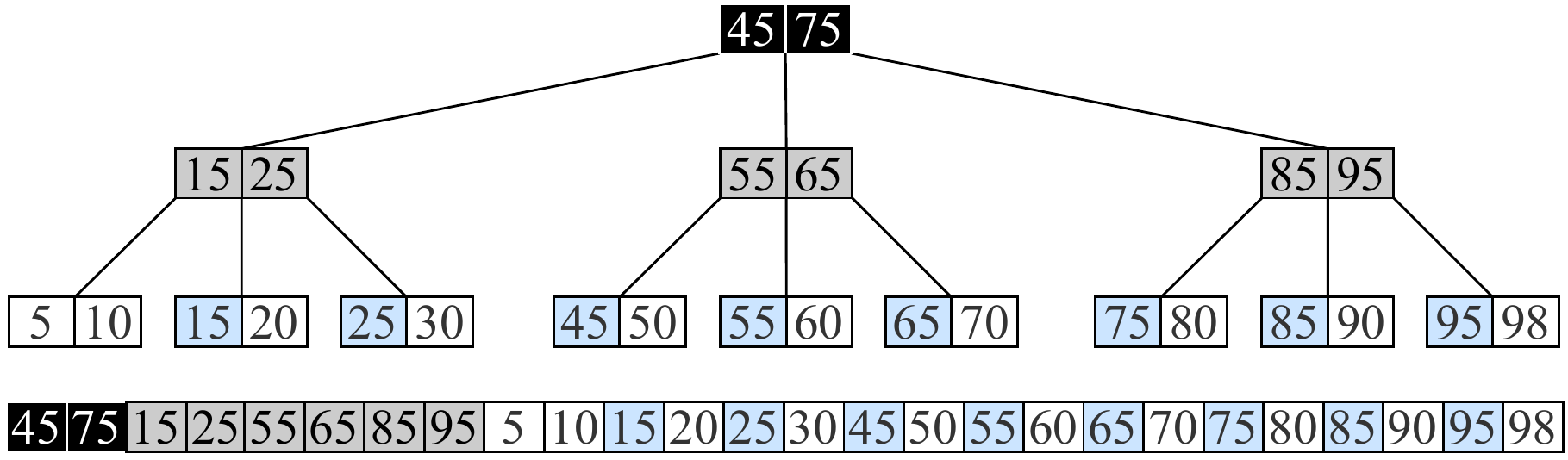}
  \vspace{-2mm}
  \caption{An example of a 3-ary linearized $B^+$-tree.
  }
  \label{fig:design-b+-tree}
  \vspace{-2mm}
\end{figure}

\textbf{Linearized $B^+$-Tree.}
\sys proposes linearized $B^+$-tree to overcome this limitation.
This variant stores all actual keys reside in the leaf nodes.
The internal nodes contain only discriminating keys that guide the search toward the target leaf.
These keys correspond to the start and end addresses of elementary intervals that are transformed from the original prefixes.
The tree also stores all keys in a single flat array,
where the internal nodes are laid out first in breadth-first order,
followed by all the leaf nodes.
As shown in the blue box in Fig. \ref{fig:design-b+-tree},
leaf nodes replicate the keys from internal nodes.

\textbf{Search on Linearized $B^+$-Tree.}
The search algorithm traverses the structure from the root to a leaf node within the flat array, as detailed in Algorithm \ref{alg:planb_search}.
%
In the search process,
the parent-child relationship in the linearized $B^+$-Tree can be computed arithmetically from array indices.
To locate the keys of a child node during traversal of the internal nodes,
we use its parent's level $d$ (with the root at $d = 0$) and the parent's 0-indexed position $j$ within that level.
The calculation involves two steps:


\begin{itemize}[
        topsep=2pt,itemsep=2pt,partopsep=0pt, parsep=0pt,leftmargin=10pt
    ]
\item Find the starting index of the child's level.
      The total number of keys in all internal node levels preceding the child's level ($d + 1$) is the sum of a geometric series.
      With $k$ keys per node and a tree arity of $b$ ($= k + 1$),
      the starting index of level $d + 1$:
      \vspace{-2mm}
      \[
          \text{level\_start}(d + 1) = k \times \frac{b^{(d+1)} - 1}{b - 1}
      \]

   \item Find the child's offset within that level.
      Each node has $b$ children, so the $j$-th node at level $d$ has its children starting at offset $j \times b$ within level $d + 1$.
      Therefore,
      the starting index of the keys for a specific child $c$ (where $c$ is from 0 to $b-1$) of the $j$-th node at level $d$ is given by:
      \vspace{-2mm}
      \[
          \text{child\_start}(d, j, c) = \text{level\_start}(d+1) + (j \times b + c) \times k
      \]
\end{itemize}

\textbf{Parameter Tuning.}
We tune $b$ and $k$ for cache efficiency,
sizing each node to fit a 64-byte cache line (common on x86 CPUs).
\sys uses 64-bit keys (most significant bits of 128-bit interval starts),
allowing $k=8$ sorted keys per node.
This yields a $B^+$-tree of order 9 (8 keys, $b=9$ children),
forming a 9-ary tree where nodes align perfectly with cache lines.
For instance,
the root node's keys occupy indices 0-7 of the array,
and the keys of its nine children are stored contiguously starting from index 8.
The design is generalizable: an architecture with 128-byte cache lines,
such as Apple Silicon \cite{Apple-Silicon},
would use 16-key nodes to form a 17-ary tree.



\re{The search on the linearized $B^+$-tree always proceeds down to a leaf node,
making the lookup process deterministic.
A tree with $d$ internal levels has a fixed search depth of $d + 1$,
meaning the worst-case lookup cost is strictly bounded by the tree's height.
This property is crucial for minimizing tail latency by avoiding the performance outliers common in trie- or hash-based approaches.}

The tree's capacity scales with its height.
A $B^+$-tree with $d$ levels of internal nodes can have up to $b^d$ leaf nodes.
With each leaf holding $k$ keys, the total capacity is $k \times b^d$.
For example, a tree with 5 non-leaf levels
has $\sum_{i=0}^{4} 9^i = 6,560$ internal nodes,
indexing $9^5 = 59,049$ leaf nodes.
The number of internal nodes is approximately a fraction of $1/(b-1) = 1/8 = 12.5\%$ of the number of leaf nodes.
With 8 keys per leaf,
the structure supports more than 472,392 elementary intervals,
which meets the needs of current FIBs.
In this configuration,
the leaf nodes ($M_{leaf}$) occupy $\sim$3.6MB of memory for 8-byte keys,
and the memory footprint of the internal nodes is $M_{internal} \approx M_{leaf} / (b-1) \approx 0.45$MB.
Consequently,
each FIB entry consumes an average of 18 bytes,
comprising 16 bytes for the leaf node and 2 bytes for internal node overhead.
\re{A tree with 6 internal levels can index over 4.2 million elementary intervals,
sufficient for future FIBs with over 1M prefixes.
For this larger structure, each lookup completes in a predictable 7 accesses.}
In this case,
the memory consumption increases to $\sim$32MB for the leaf nodes
and 4MB for the internal nodes.
The total size of $\sim$36MB remains well within the L3 cache available on modern server-class CPUs.

\setlength{\textfloatsep}{10pt}
\begin{algorithm}[t]
\footnotesize
\caption{Search on Linearized $B^+$-tree}
\label{alg:planb_search}
\begin{algorithmic}[1]
\State \textbf{Input:} \texttt{prefix}: The target IPv6 prefix. \texttt{key[]}: The flat array representing the linearized $B^+$-tree. \texttt{d}: The depth of internal nodes in the tree. \NoNumber{\texttt{k}: keys per node.}
\State \textbf{Output:} \texttt{final\_idx}: The index in the leaf-node region of \texttt{key[]} for the matching elementary interval.
\vspace{1mm}
\State b $\gets$ k $+ 1$
\State node\_off\_in\_level $\gets$ 0

\For{i $\gets$ 0 \textbf{to} $d-1$}
    \State level\_start $\gets$ k $\times$ $\left(\frac{b^{i} - 1}{b - 1}\right)$
    \State node\_start $\gets$ level\_start $+$ node\_off\_in\_level $\times$ k
    \State node\_idx $\gets$ \texttt{lower\_bound\_binary\_search}(key, node\_start, k,
    \NoNumber{prefix)}
    \State node\_off\_in\_level $\gets$ node\_off\_in\_level $\times$ b $+$ node\_idx
\EndFor

\State leaf\_start $\gets$ k $\times$ $\left(\frac{b^{d} - 1}{b - 1}\right)$
\State leaf\_node\_start $\gets$ leaf\_start $+$ node\_off\_in\_level $\times$ k
\State final\_idx $\gets$ \texttt{lower\_bound\_binary\_search}(key,
\NoNumber{leaf\_node\_start, k, prefix) + leaf\_node\_start}

\State \Return final\_idx
\end{algorithmic}
\end{algorithm}

\textbf{Maintaining Structural Integrity.}
To guarantee efficient arithmetic indexing and search,
\sys maintains the structural integrity of a complete b-ary tree.
This is achieved by mandating that all nodes have a fixed, full layout.
All nodes on non-leaf levels must be full;
partially filled nodes are padded with sentinel values (i.e., 0xFFFFFFFFFFFFFFFF) to achieve this.
For example, if a node with eight key slots has only five valid entries,
the remaining three are filled with sentinels.
Leaf nodes are also padded,
but \sys applies a memory optimization by concentrating sentinel values in the rightmost nodes of each level in a bottom-up manner.
This approach eliminates the need to allocate memory space for trailing leaf nodes that would otherwise contain only sentinels.
Although internal nodes consisting entirely of sentinel keys still occupy memory,
they are excluded from the search process and are never accessed,
thereby avoiding CPU cache loads.
Padding preserves the property that child locations can be derived solely through index arithmetic,
eliminating the need for additional metadata such as node sizes or linked pointers.
In practice,
\sys uses a sentinel key value for padding,
which guarantees correct search behavior while preventing misrouting.


\textbf{Separation of Keys and Values.}
The $B^+$-tree itself is represented in
a single, dense array containing only the sorted 64-bit interval start addresses (the keys).
The corresponding next-hop information (the values) is stored in a separate array.
The two arrays are associated implicitly by their indices:
the value corresponding to the key at index $i$ in the leaf-node region of the key array is located at $values[i - leaf\_start]$,
where $leaf\_start$ denotes the first index of the leaf nodes in the key array.
During a lookup,
the traversal algorithm only needs to access the key array to find the correct index.
Once the search completes and the index is identified,
the next-hop is retrieved in a single memory access.

\begin{listing}[t]
\centering
\small
\caption{Vectorized Search on Linearized B$^+$-tree.}
\vspace{-1mm}
\begin{minted}[linenos,xleftmargin=12pt,numbersep=8pt,fontsize=\footnotesize,style=default]{cpp}
// k: keys per node, key: B+-tree keys array
long LBTree::vectorized_search(uint64_t prefix) {
    long i = 0,c = -1, d = tree_depth;
    __m512i vx = _mm512_set1_epi64(prefix);
    do {
        i = (k + 1) * i + (c + 1) * k;
        __m512i node = _mm512_load_si512(&key[i]);
        c = _mm512_cmpge_epu64_mask(vx, node);
        c = __builtin_popcount(c);
    } while (--d);
    return i+c;
}
\end{minted}
\label{lst:cpp-vector-search}
\vspace{-1mm}
\end{listing}

\vspace{-2mm}
\subsection{Vectorization and SIMD Acceleration}
\label{sec:vectorization}

\sys exploits data-level parallelism through modern CPU SIMD extensions,
particularly AVX-512 \cite{Intel-AVX512}.
\re{The proposed design is portable across diverse SIMD architectures,
such as AVX2 \cite{Intel-AVX512}, SSE \cite{Intel-SSE4}, and ARM NEON \cite{ARM-Neon}.}
The contiguous, cache-line-aligned layout of our linearized $B^+$-tree is critical for enabling efficient SIMD operations.
Instead of performing a sequence of scalar comparisons,
\sys's search algorithm executes the in-node search using a few highly efficient vector instructions.
Listing \ref{lst:cpp-vector-search} shows this algorithm,
and Listing \ref{lst:asm-vector-search} shows the corresponding assembly code generated by the compiler and disassembled using \texttt{gdb}.
The process for a node containing 8 keys on an AVX-512 capable CPU is as follows:

\begin{enumerate}[
        label=(\roman*), 
        topsep=2pt,itemsep=2pt,partopsep=0pt,parsep=0pt
    ]

    \item \textbf{Broadcast.} The 64-bit target address is broadcast into all 8 slots of a 512-bit vector register (line 4, Listing \ref{lst:cpp-vector-search}).
    
    \item \textbf{Vector Load.} The 8 sorted 64-bit keys of a $B^+$-tree node are loaded from memory into a second 512-bit SIMD register (line 7, Listing \ref{lst:cpp-vector-search}).
  
    \item \textbf{Parallel Comparison.} A single SIMD instruction is executed to compare the target address against all 8 keys in the node simultaneously (line 8, Listing \ref{lst:cpp-vector-search}).
    Certain compilers can fuse this operation with the preceding load, generating a single assembly instruction such as \texttt{vpcmpnltuq}  (line 8, Listing \ref{lst:asm-vector-search}).
    \item \textbf{Child Traversal.} The comparison results are used to determine the index of the next child node to visit (line 9 \& 6 in Listing \ref{lst:cpp-vector-search}). The process then repeats from Step (ii) at the next tree level until a leaf node is reached.
  
\end{enumerate}

    
    
    



\begin{listing}[t]
\centering
\footnotesize
\caption{Assembly Code Snippet of The Vectorized Search.}
\vspace{-1mm}
\begin{minted}[linenos,xleftmargin=12pt,numbersep=8pt,fontsize=\footnotesize,style=xcode,highlightlines=none]{c}
<+00>: vpbroadcastq %rsi, %zmm0
<+06>: movl   0x8(%rdi), %ecx
<+09>: movq   (%rdi), %rsi
<+12>: orq    $-1, %rdx
<+16>: xorl   %eax, %eax
<+18>: leaq   (%rax,%rax,8), %rax
<+22>: leaq   0x8(%rax,%rdx,8), %rax
<+27>: vpcmpnltuq (%rsi,%rax,8), %zmm0, %k0
<+35>: kmovb  %k0, %edx
<+39>: popcntl %edx, %edx
<+43>: decq   %rcx
<+46>: jne    <+18>
<+48>: addq   %rdx, %rax
<+51>: retq
\end{minted}
\label{lst:asm-vector-search}
\end{listing}

By replacing a loop of scalar operations with a short sequence of powerful vector instructions,
this SIMD-accelerated approach dramatically reduces the number of CPU cycles required to traverse each node in the $B^+$-tree.
For example,
in a $B^+$-tree of height 7,
a lookup requires only seven vector loads and comparisons,
substantially fewer than the dozens of scalar comparisons performed in a conventional binary search.
\re{On platforms with different SIMD widths,
like AVX2 (256-bit) or ARM NEON (128-bit),
the principle remains identical.
The in-node search is simply composed of multiple vector operations to cover all keys in the node (e.g., two 256-bit operations for an 8-key node),
demonstrating the flexibility and hardware-agnostic nature of our approach.}

\vspace{-2mm}
\subsection{Batching}
\label{sec:batching}

\sys processes lookups in batches rather than handling packets one by one.
Processing addresses individually underutilizes the CPU,
as the unavoidable latency of key instructions forces the pipeline to wait.
Batching mitigates this by enabling the processor to work on multiple lookups in parallel,
hiding latency and maximizing the use of its execution units.

First,
the complex SIMD instructions central to \sys's node search,
such as \texttt{vpcmpnltuq} (line 8, Listing \ref{lst:asm-vector-search}),
exhibit multi-cycle latency.
In a single-lookup model,
the CPU would issue the comparison and then stall,
waiting for the results before it could execute the subsequent instruction.
By processing a batch of lookups,
\sys provides the CPU with independent work.
This pipelined execution of instructions from different lookups keeps the processor's resources constantly engaged,
effectively hiding instruction latency and sustaining a high rate of completed operations.

Second,
after a SIMD comparison,
the intermediate results must be transferred from vector units into general-purpose units before determining the correct child index.
Although efficient,
this transfer still introduces overhead.
By treating a bundle of addresses as a batch,
\sys amortizes this per-lookup cost,
executing transfer and post-processing instructions in a way that maximizes throughput across the entire batch rather than paying the cost repeatedly for each lookup.

\re{Our DPDK implementation (Section \ref{sec:system}) forms batches directly from the bursts of packets retrieved from the NIC's RX ring.
This approach maximizes throughput by processing available packets immediately,
up to the hardware vector width,
without incurring any artificial latency for batch formation.
Batching transforms the lookup from a series of latency bound, sequential operations into a highly parallel, throughput oriented pipeline.
Such a design synergizes with our other optimizations to fully exploit the instruction level parallelism of modern CPUs.}

\vspace{-3mm}
\subsection{Branch-Free Traversal and Loop Unrolling}
\label{sec:branch-free}



    
    
    
    



A standard $B^+$-tree traversal involves a sequence of comparisons within each node to find the correct child branch.
Implemented naively with if-else statements,
this process creates conditional branches in the instruction stream.
For unpredictable input traffic,
these branches are highly susceptible to misprediction by the CPU's branch predictor.
A single misprediction incurs a significant performance penalty,
forcing the CPU to flush its speculative execution pipeline and restart from the correct path,
which can cost dozens of cycles and severely degrade lookup throughput.
\sys's traversal algorithm is designed to be entirely branch-free.
Instead of conditional jumps,
we employ specialized CPU instructions that manipulate data
based on comparison outcomes without altering the control flow.
We leverage two key types of branch-free instructions:
vectorized comparisons, which produce a bitmask,
and conditional moves.

First, we can replace a sequence of scalar comparisons with a single vectorized comparison as described in Section \ref{sec:vectorization}.
This operation compares the target address against all keys in a node simultaneously using an SIMD instruction.
The result of this parallel comparison is an 8-bit mask,
where each bit corresponds to the outcome of one of the 64-bit lane comparisons.
%
Next,
we apply a population count (\texttt{popcnt}) instruction to this mask (line 9 in Listing \ref{lst:cpp-vector-search} and line 10 in Listing \ref{lst:asm-vector-search}),
The resulting count directly provides the index of the child pointer to traverse next.
This single instruction replaces an entire loop of bit checks,
making the process of finding the correct branch from the comparison result extremely fast and branch-free.

Alternatively,
in scenarios where vector instructions are not used,
the target address is compared against a node's keys,
setting flags in the CPU's status register.
\sys then use these flags to conditionally update the index to the next node without executing a jump.
Both techniques are integral
to \sys's design,
as they replace control-flow dependencies with data-flow dependencies,
which modern out-of-order CPUs can execute far more efficiently.

To further optimize performance,
we fully unroll the nested traversal loops (per-tree-level search $\times$ per-batch lookup).
Each iteration consists of only a handful of instructions,
including one vector comparison, one move-mask,
a single \texttt{ctz},
and an arithmetic calculation for indexing the child node.
Because the linearized $B^+$-tree is shallow (seven levels are sufficient to index over one million intervals)
and lookup batches are typically small (tens of addresses per core),
the total number of operations per lookup remains bounded and predictable.
This makes complete unrolling feasible at compile time without code-size explosion or instruction-cache pressure.
The resulting straight-line instruction sequence eliminates all loop-control overhead and maximizes the compiler's ability to reorder instructions aggressively,
hide instruction latencies, and issue operations in parallel.



\vspace{-3mm}
\subsection{Dynamic FIB Update}
\label{sec:update}

\begin{algorithm}[t]
\footnotesize
\caption{Linearized $B^+$-tree Construction from FIB}
\label{alg:build_planb}
\begin{algorithmic}[1]
    \State \textbf{Input:} A FIB $F = \{f_1, f_2,...\}$. Each entry $f$ is $(f.\text{prefix}, \ f.\text{length}, \ f.\text{next\_hop})$.
    \State \textbf{Tree parameters:} depths $d$, keys per node $k$, arity $b=k+1$.
    \State \textbf{Output:} Linearized $B^+$-tree of keys array $key$ and next-hop array $value$. 
    \State Let $E$ be an empty list of tuples $(address, next\_hop)$.
    \State $E \leftarrow \{(f.\text{prefix}, \ f.\text{next\_hop}), (f.\text{prefix} + 2^{64 - f.\text{length}}, \ -1), ...
    | \ f \in F\}$
    \State Sort $E$ lexicographically by $(address, length)$
    \State Let $S$ be an empty stack for next-hops
    \For{each element $e \in E$}
        \If{$e.\text{next\_hop} \ge 0$}
            \State $S.\text{push}(e.\text{next\_hop})$
        \Else
            \State $S$.pop()
            \If{$S$ is not empty}
                \State $e.\text{next\_hop} \leftarrow S.\text{top()}$
            \EndIf
        \EndIf
    \EndFor

    \State $L_{\text{start}} \leftarrow level\_start(d-1)$
    \For{$i \leftarrow 0$ to $|E|-1$}
        \State $key[L_{\text{start}} + i] \leftarrow E[i].\text{prefix}$
        \State $value[i] \leftarrow E[i].\text{next\_hop}$
    \EndFor
    \State Fill remaining entries in the leaf level with sentinels.


    \For{$l \leftarrow d-1$ to $1$}
        \State $p_{\text{start}} \leftarrow level\_start(l-1)$
        \State $c_{\text{start}} \leftarrow level\_start(l)$
        \For{$i \leftarrow 0$ to $b^{l}-1$ in steps of $k \times (k + 1)$}
            \For{$j \leftarrow 1$ to $k$}
                \State $key[p_{\text{start}}] \leftarrow key[c_{\text{start}} + i + k \times j]$
                \State $p_{\text{start}} \leftarrow p_{\text{start}} + 1$
            \EndFor
        \EndFor
    \EndFor
    \State \Return $(key, value)$
\end{algorithmic}
\end{algorithm}

Handling dynamic FIB updates without compromising lookup performance is a critical requirement for practical routing systems.
\sys adopts a batch-oriented, rebuild-and-swap model that cleanly separates updates from lookups.
The update process is triggered for a batch of one or more prefix changes (insertions, deletions, or modifications).
Instead of altering the live data structure,
\sys performs the update in the background by following these steps:

\begin{itemize}[
        topsep=2pt,itemsep=2pt,partopsep=0pt, parsep=0pt,leftmargin=10pt
    ]
  \item \textbf{Batch Modify.} Updates are accumulated into a secondary prefix list maintained in a seperate memory region, preferably by other CPU cores.
  Multiple pending changes can be coalesced into a single batch.

  \item \textbf{Complete Rebuild.} The entire pre-computation process, including range conversion, interval partitioning, and linearized $B^+$-tree construction, is executed on the prefix list as detailed in Algorithm \ref{alg:build_planb}.
  The result is a brand-new lookup structure that is compact and consistent.
  
  \item \textbf{Atomic Swap.}
  Once fully built, the system performs a single atomic pointer update to redirect future lookups to the new structure. Threads in the middle of a lookup continue using the old structure until they complete, after which it can be safely deallocated.
\end{itemize}

This strategy provides several benefits.
\re{First, the lookup path remains strictly contention-free.
Since lookups continue to operate on the stable,
existing data structure while the new one is built in the background,
they suffer no performance degradation or stalls from update activity.
No locks or other synchronization primitives are needed in the forwarding fast path.
Second, the transition is seamless.
Once the new structure is ready, a single atomic pointer swap redirects all new lookups.
This guarantees that every lookup operation completes correctly without being affected by the update.
Third, by batching updates and performing a full rebuild, \sys amortizes the update cost and efficiently handles high-volume or bursty route changes. This makes its performance predictable and robust, regardless of the update pattern.}

\re{The trade-off is a temporary increase in memory usage to hold both the old and new structures.
However, this overhead is significantly mitigated by \sys's inherent compactness.
\sys's lean design ensures that even during an update,
the total memory consumption is less likely to evict other critical data from the CPU cache.
Once all in-flight lookups on the old structure complete,
it is safely deallocated and has no further impact on the system.}

\vspace{-3mm}
\subsection{Implementation}
\label{sec:system}

We implemented a prototype software router for \sys using the Data Plane Development Kit (DPDK) \cite{DPDK-PerformanceThread,Metronome-ToN}.
Our implementation leverages thread specialization and batch processing pipelines to meet three key requirements:
(i) fully exploiting \sys's SIMD-accelerated search,
(ii) sustaining line-rate forwarding with large, realistic IPv6 FIBs,
while being robust to anticipated growth in FIB size and complexity,
and (iii) scaling efficiently across multi-core processors with minimal coordination overhead.

\begin{figure}[t]
  \centering
  \includegraphics[width=0.99\linewidth]{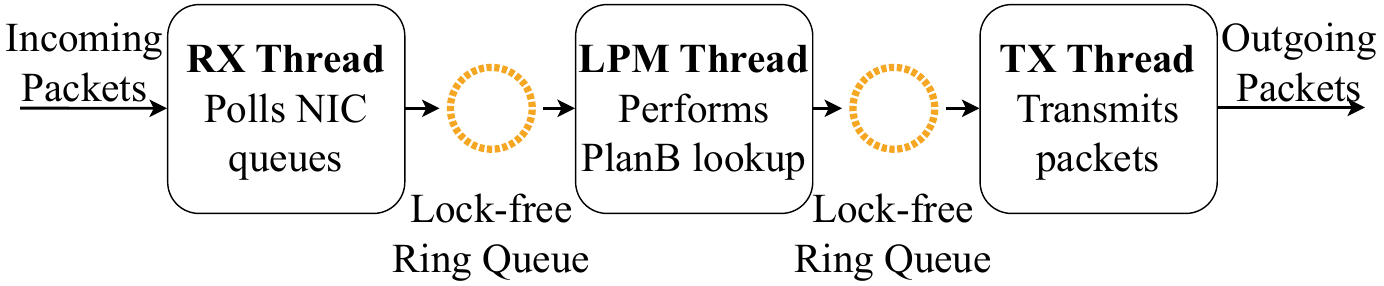}
  \vspace{-2mm}
  \caption{\sys's RX-LPM-TX module within DPDK.
  }
  \label{fig:design_dpdk_mod}
  \vspace{-2mm}
\end{figure}

\textbf{Thread Pipeline.}
To prevent other packet processing tasks from degrading LPM cache hit rate,
we extend the DPDK performance-thread RX-TX model \cite{DPDK-PerformanceThread} by introducing a dedicated lookup stage between packet reception and transmission.
In this RX-LPM-TX model (illustrated in Fig. \ref{fig:design_dpdk_mod}),
RX threads poll NIC queues and enqueue bursts of packets into per-queue software rings.
Lookup (LPM) threads dequeue those bursts,
execute \sys's $B^+$-tree search to resolve next-hop indices,
and annotate packets with egress metadata.
Finally, TX threads batch, format, and transmit the annotated packets to the appropriate NIC port.
This separation enables independent scaling of each stage:
RX threads scale with NIC queues,
LPM threads scale with the cost of IPv6 lookups,
and TX threads scale with output load.

\textbf{Cache Residency.}
Modern multi-core CPUs typically feature private L1/L2 caches for each core and a shared L3 cache.
This L3 cache can be either fully shared across all cores or partitioned by groups of cores (e.g., the CCDs/CCXs in AMD Zen5 architecture).
For architectures with such group-shared L3 caches,
\sys leverages the hardware topology to ensure cache isolation.
It pins the LPM threads to cores in one cache group, while confining the RX/TX and management threads to cores in a separate group.
This strategy prevents the LPM thread's L3 cache slice from being evicted by other system activities,
thereby preserving its cache residency and maintaining high, predictable performance.

\textbf{Design Trade-offs.}
A key trade-off of our three-stage RX-LPM-TX pipeline is a slight increase in per-packet latency over a conventional RX-TX model.
However, for IPv6 workloads where lookup cost dominates,
this overhead is offset by reduced contention, improved SIMD batching, and better load distribution.
For less demanding workloads,
such as those with smaller IPv4 or IPv6 FIBs,
\sys supports a fused two-stage RX-(LPM+TX) configuration.
In this model, the lookup and transmission logic are combined into a single thread,
eliminating the extra ring crossing while still leveraging cache-friendly, SIMD-accelerated lookups.
Moreover, the number of LPM+TX threads can be tuned independently of the RX threads.
This flexibility allows \sys to be optimized for different scenarios:
a three-stage pipeline for lookup-bound workloads
and a two-stage pipeline for I/O-bound ones.




\vspace{-3mm}
\section{Evaluation}
\label{sec:evaluation}
\vspace{-2mm}

We have implemented \sys in C++ as a user-space application on Linux.
Our evaluation is conducted on two hardware platforms.
We first employ two servers,
each featuring a 24-core Intel Xeon 8331C processor (3GHz turbo, 36MB L3 cache) and 1TB of DDR5 RAM.
The servers are connected to a 400Gbps Ethernet switch
and runs Ubuntu 22.04 with Linux kernel 5.15.0-143-generic.
The second platform is a commodity system equipped with an AMD Ryzen 9 370HX mobile processor.
The processor features performance cores (Zen5, up to 5.1GHz) and efficiency cores (Zen5c, up to 3.3GHz).
The system is configured with 128GB DDR5 RAM and runs the same OS and kernel version as the server.

\textbf{Evaluated Schemes.}
We compare \sys against \re{four} state-of-the-art software-based IP lookup schemes: PopTrie \cite{Poptrie}, \re{CP-Trie \cite{CP-Trie}, Neurotrie \cite{Neurotrie-ToN} (specifically the Neurotrie-S variant)} and HBS \cite{HBS-ToN}.
The schemes are implemented in C++, integrated with DPDK for packet I/O, and evaluated using the same environment.
To ensure fairness, we use a uniform DPDK version, identical DPDK configurations, and compile all code with g++ 11.4.0 using the -O3 optimization flag on both platforms.
All evaluated results are averaged over multiple runs to ensure statistical significance.

\textbf{Datasets.}
We use rrc00-19--rrc00-25 in Table \ref{table:datasets} to test,
as the prefix distribution and numbers of rv1--rv7 is similar to rrc00-25.
To assess scalability,
we generate four synthetic FIBs (synth-1 to synth-4) with 250K, 500K, 750K, and 1M prefixes, respectively.
The prefix length distribution of these synthetic FIBs is modeled after the rrc00-25 dataset to ensure realism.
For evaluating lookup performance,
we generate a test trace for each FIB.
Each trace contains a number of lookups equal to 100 times the number of prefixes in the corresponding FIB. These lookups are generated by randomly sampling prefixes from the FIB
and then creating a destination IP address that matches each sampled prefix.
For update evaluation,
we collecte one-hour update traces for rrc00 between 8:00 AM and 9:00 AM for every year in the dataset \cite{RIPE-RIS}.
Moreover, in the trace-driven benchmarks,
we observe that loading trace records (64-bit prefixes) from memory significantly degrades overall performance.
To address this, we prefetch 
these records into spare vector registers, which will be 
reused across subsequent iterations.

\textbf{Performance Metrics.}
We evaluate \sys and other systems across the following key performance dimensions:

\begin{itemize}[
        topsep=2pt,itemsep=2pt,partopsep=0pt, parsep=0pt,leftmargin=10pt
    ]
    \item System Throughput: The end-to-end forwarding rate of the DPDK-based L3 application for 64-byte packets, measured in Million Packets Per Second (MPPS).
  \item Lookup Speed: The raw performance of the IP lookup module, measured as a microbenchmark. We report this in MLPS and BLPS.
  
  \item Memory Overhead: The total memory consumed by the lookup data structure, measured in Megabytes (MB).
  
  \item Update Overhead: The time required to perform batch updates of prefixes, measured in microseconds (us).
\end{itemize}

\re{To understand the contributions of various design choices,
we further conduct an ablation study
by selectively disabling key designs in \sys and measuring the resulting performance impact.}
Each experiment are repeated 20 times to ensure reliability, with the average value of each metric are reported.

\begin{figure}[t]
    \center
    \includegraphics[width=0.99\linewidth]{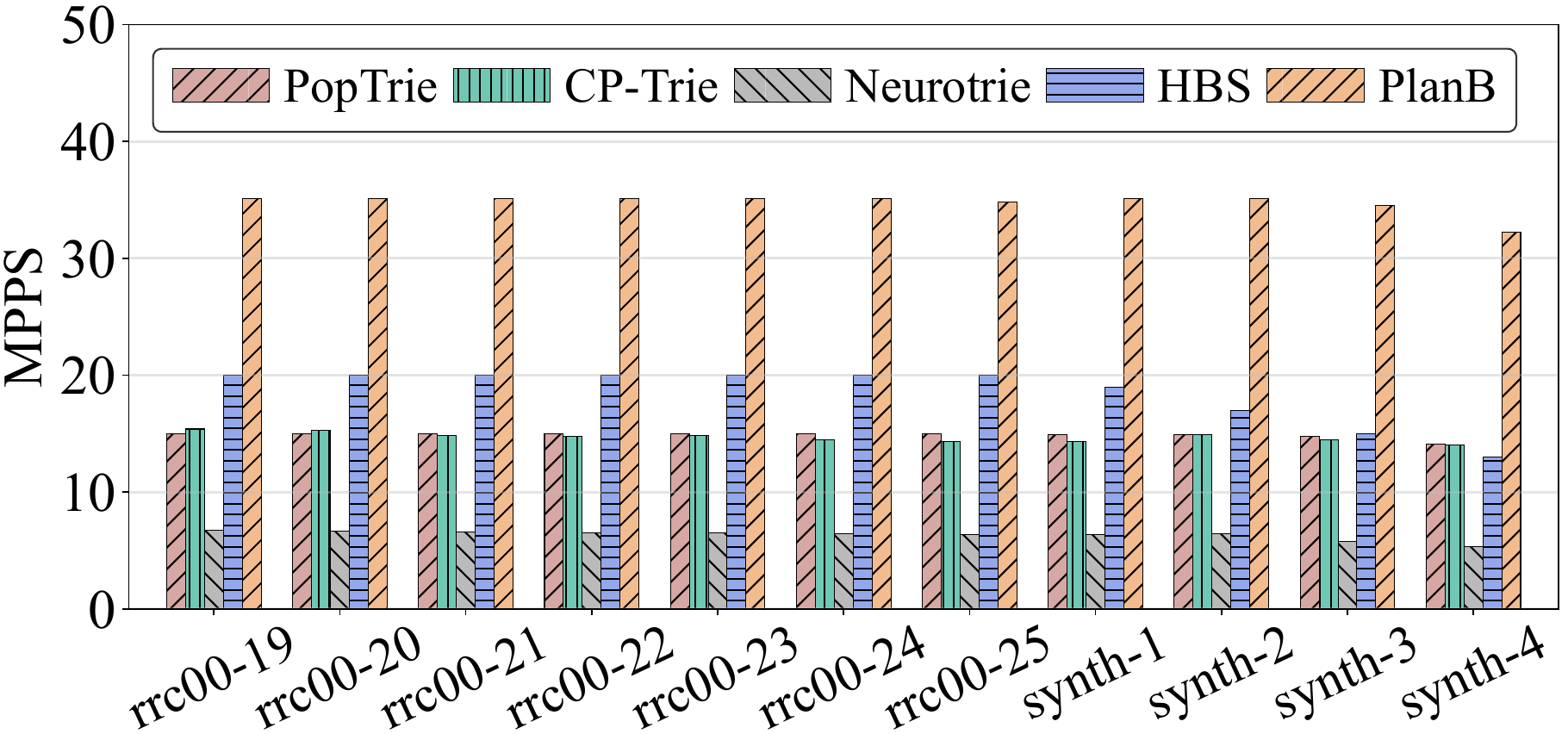}
    \vspace{-3pt}
    \caption{\label{fig:2_eva_throughput_amd} System throughput for DPDK-based L3 forwarding.
    }
    \vspace{-7pt}
\end{figure}

\begin{figure}[t]
    \center
    \includegraphics[width=0.99\linewidth]{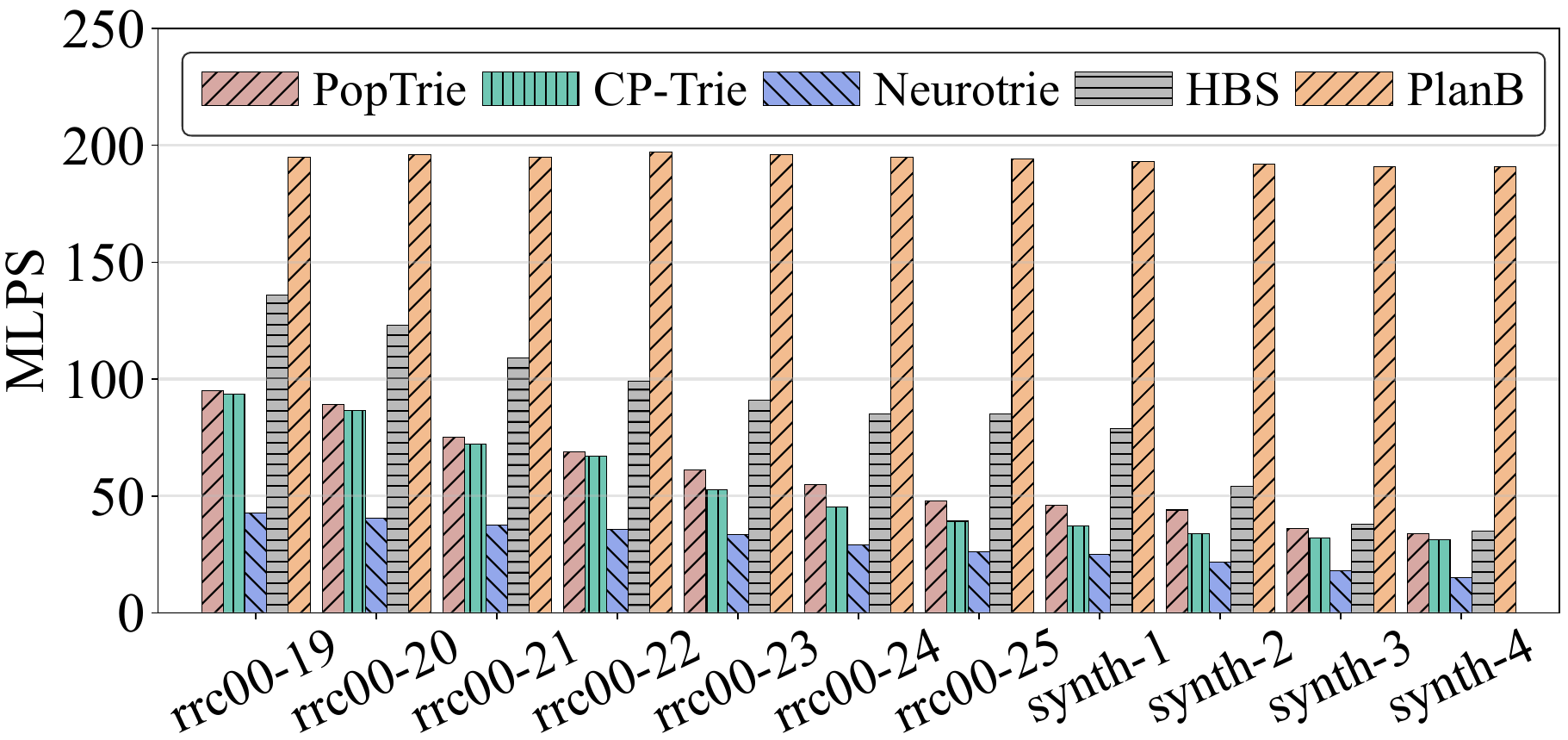}
    \vspace{-3pt}
    \caption{\label{fig:eva_mlps_intel_single_core} Lookup speed with single core on Intel CPU.
        }
    \vspace{-7pt}
\end{figure}

\begin{figure}[t]
    \center
    \includegraphics[width=0.99\linewidth]{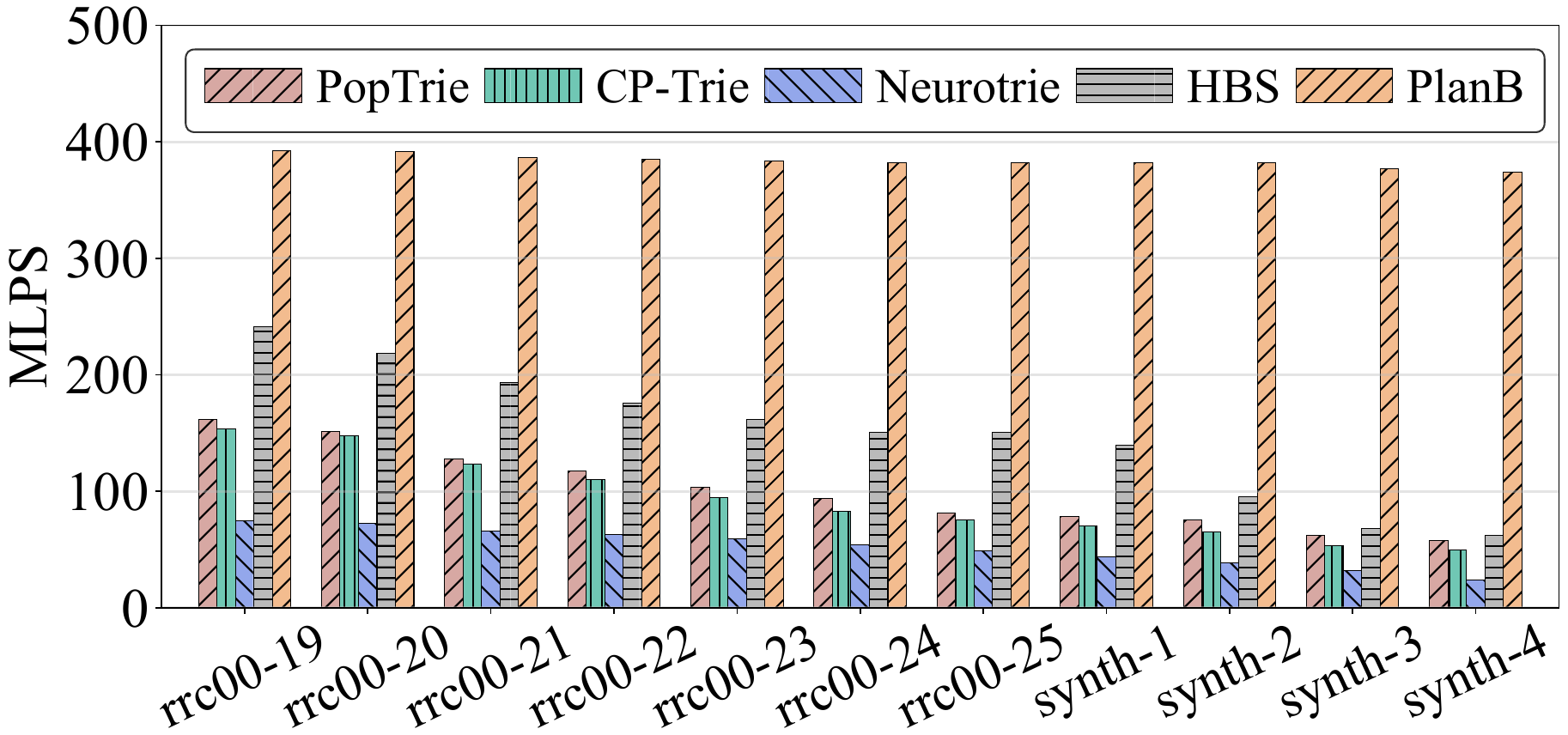}
    \vspace{-3pt}
    \caption{\label{fig:eva_mlps_amd_single_core} Lookup speed with single core on AMD CPU.
        }
\end{figure}


\begin{figure*}[tb]
    \begin{minipage}[t]{0.489\textwidth}
        \vfill
        \includegraphics[width=\linewidth]{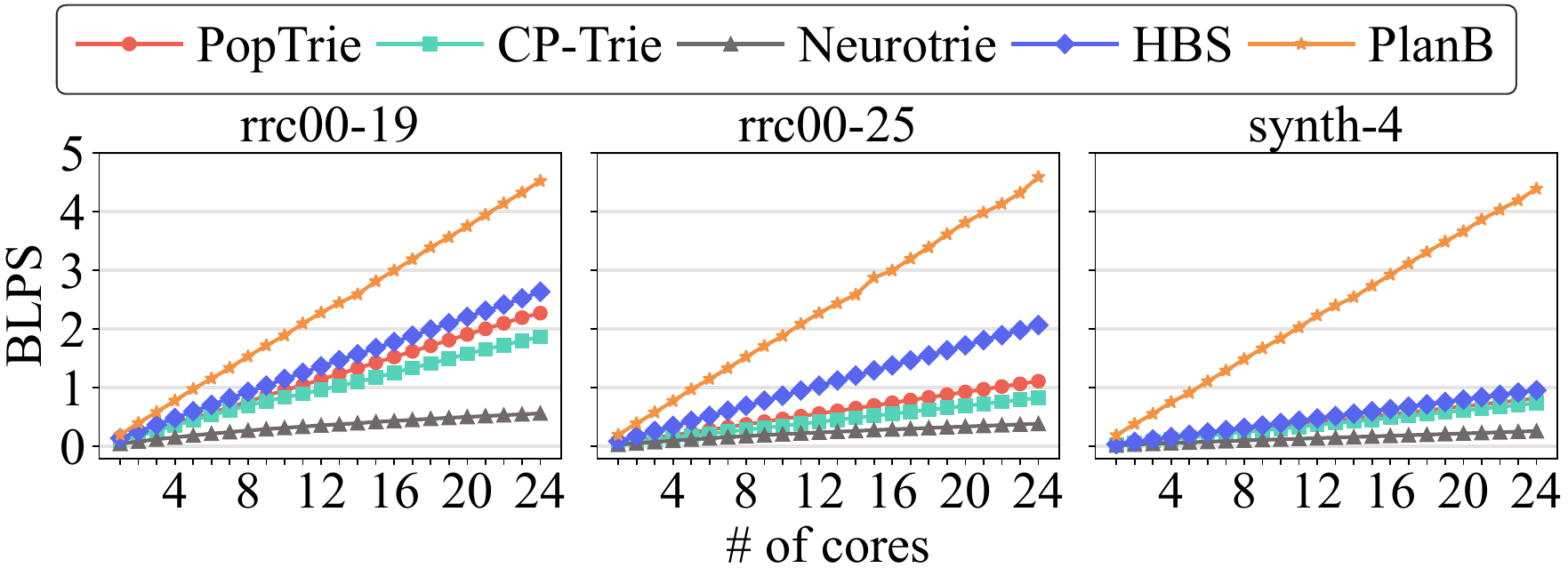}
        \vspace{-15pt}
        \caption{\label{fig:eva_mlps_intel_multi} Lookup speed with multi-cores on Intel CPU.
        }
    \end{minipage}
    \hfill
    \begin{minipage}[t]{0.489\textwidth}
        \vfill
        \includegraphics[width=\linewidth]{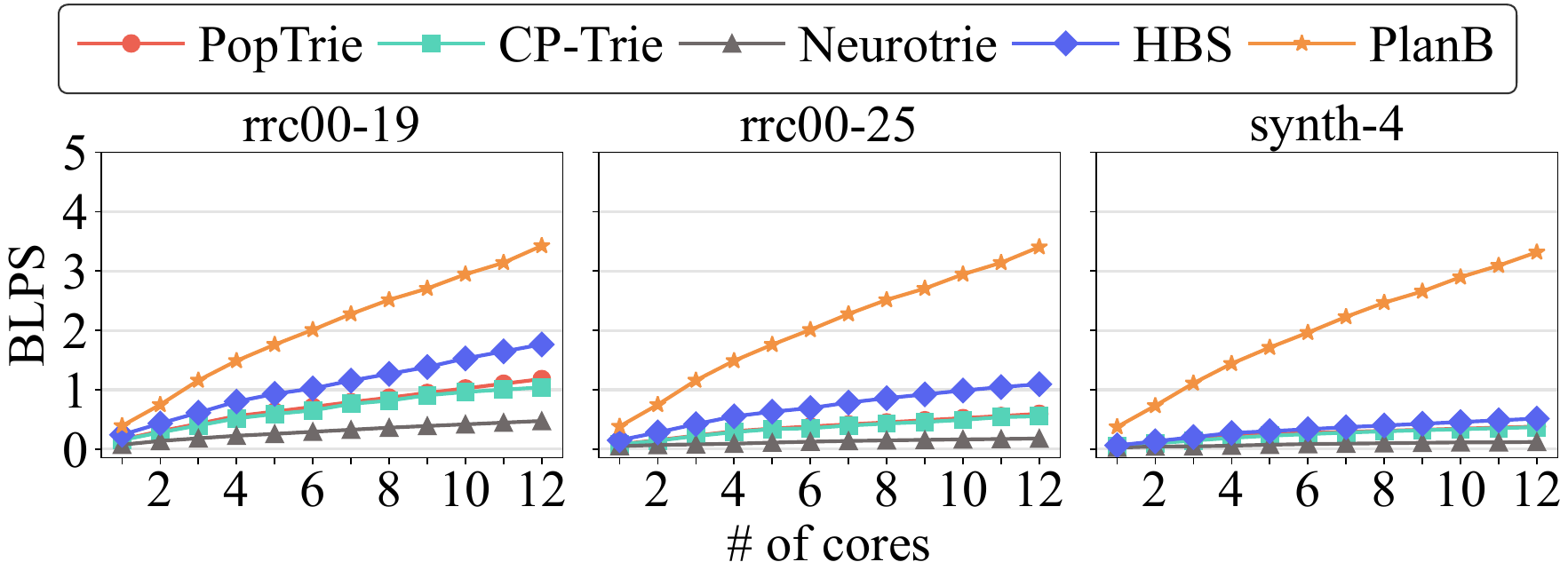}
        \vspace{-15pt}
        \caption{\label{fig:eva_mlps_amd_multi} Lookup speed with multi-cores on AMD CPU.
        }
    \end{minipage}
    \hfill
    \vspace{-10pt}
\end{figure*}

\vspace{-2mm}
\subsection{System Throughput}
\label{sec:system-throughput}
\vspace{-1mm}

We first evaluate the end-to-end system throughput using \sys-enabled DPDK
for back-to-back 64-byte packet forwarding.
Fig. \ref{fig:2_eva_throughput_amd} shows the results.
\re{Across all evaluated FIBs, \sys achieves a system throughput
that is 2.3$\times$$\sim$2.8$\times$ faster than PopTrie,
2.3$\times$$\sim$2.9$\times$ faster than CP-Trie, 4.7$\times$$\sim$5.8$\times$ faster than Neurotrie,
and 1.7$\times$$\sim$2.5$\times$ faster than HBS.}
As the number of prefixes in the routing tables increases,
the performance gap widens, highlighting \sys's superior scalability.
\re{Whereas trie-based schemes such as PopTrie, CP-Trie, and Neurotrie are constrained by costly memory indirection and conditional logic,}
\sys avoids these overheads by transforming the lookup into a 1D search.
This is achieved by partitioning the address space into elementary intervals and performing an efficient search on a linearized $B^+$-tree.
This dense, contiguous layout with vectorized search and branch-free logic,
enables \sys to fully leverage modern CPU architectures,
delivering high throughput even with large and complex FIBs.
Moreover,
\sys is integrated into a dedicated lookup stage between packet reception and transmission in the DPDK pipeline,
ensuring that its high lookup speed translates into tangible end-to-end forwarding performance.

\vspace{-3mm}
\subsection{Lookup Speed}
\label{sec:lookup-speed}
\vspace{-1mm}

We measure the lookup speed in MLPS/BLPS on both single-core and multi-core configurations with Intel and AMD CPUs 
to assess their absolute performance and scalability.

\textbf{Single-Core Performance.}
Figs. \ref{fig:eva_mlps_intel_single_core} and \ref{fig:eva_mlps_amd_single_core}
illustrate the single-core lookup performance across a range of real-world and synthetic IPv6 FIBs.
On the Intel server,
\sys consistently delivers a lookup speed between 191 and 197 MLPS.
\re{This performance represents a speedup of 2$\times$$\sim$5.6$\times$ over PopTrie
2.1$\times$$\sim$5.8$\times$ over CP-Trie,
4.3$\times$$\sim$9.6$\times$ over Neurotrie,
and 1.4$\times$$\sim$5.6$\times$ over HBS.}
The performance advantage of \sys is even more pronounced on the AMD mobile processor,
where it achieves a remarkable 374$\sim$393 MLPS.
\re{This represents a speedup of 2.3$\times$$\sim$6.7$\times$ over PopTrie,
2.6$\times$$\sim$7$\times$ over CP-Trie,
5.3$\times$$\sim$12.5$\times$ over Neurotrie and 1.6$\times$$\sim$6.3$\times$ over HBS.}
The significant performance improvement of \sys stems from its fundamental design.
By converting LPM to a 1D search, \sys executes a fixed, small number of highly optimized steps to find the result.
\re{The performance of trie-based schemes degrades with longer prefixes that require deeper trie traversal.
While Neurotrie attempts to mitigate this with a DRL-optimized shallow trie,
its real-world performance is bottlenecked by high per-node overhead.
Each step in its lookup requires chasing pointers between separate memory structures and executing complex scalar logic.}
HBS's performance is sensitive to the number of distinct prefix lengths and the efficiency of its hash function.

\textbf{Multi-Core Scalability.}
Figs. \ref{fig:eva_mlps_intel_multi} and \ref{fig:eva_mlps_amd_multi}
show the aggregated lookup throughput as the number of active cores increases.
\sys demonstrates near-linear scalability on both CPU architectures,
a critical attribute for modern multi-core packet processing engines.
On the 24-core Intel server, \sys's throughput scales to a peak of 4.6 BLPS.
\re{This is 2$\times$$\sim$5.6$\times$ higher than PopTrie,
2.4$\times$$\sim$6$\times$ higher than CP-Trie,
8.1$\times$$\sim$14$\times$ higher than Neurotrie and 1.7$\times$$\sim$4.8$\times$ higher than HBS at maximum core counts.}
On the 12-core AMD mobile processor,
\sys achieves up to 3.4 BLPS,
\re{outperforming PopTrie by 2.9$\times$$\sim$8.7$\times$,
CP-Trie by 3.3$\times$$\sim$9.0$\times$, Neurotrie by 7.2$\times$$\sim$18.3$\times$,
and HBS by 1.8$\times$$\sim$6.7$\times$.}
\sys's batching and branch-free logic hide instruction latencies and avoid misprediction penalties,
which plague PopTrie, CP-Trie and Neurotrie with their pointer-based,
conditional traversals and HBS's heuristic searches.
Loop unrolling further optimizes the shallow tree depth (typically 6--7 levels),
enabling linear multi-core scaling.

\vspace{-3mm}
\subsection{Memory Overhead}
\label{sec:memory-overhead}

\begin{figure}[t]
    \center
    \includegraphics[width=0.99\linewidth]{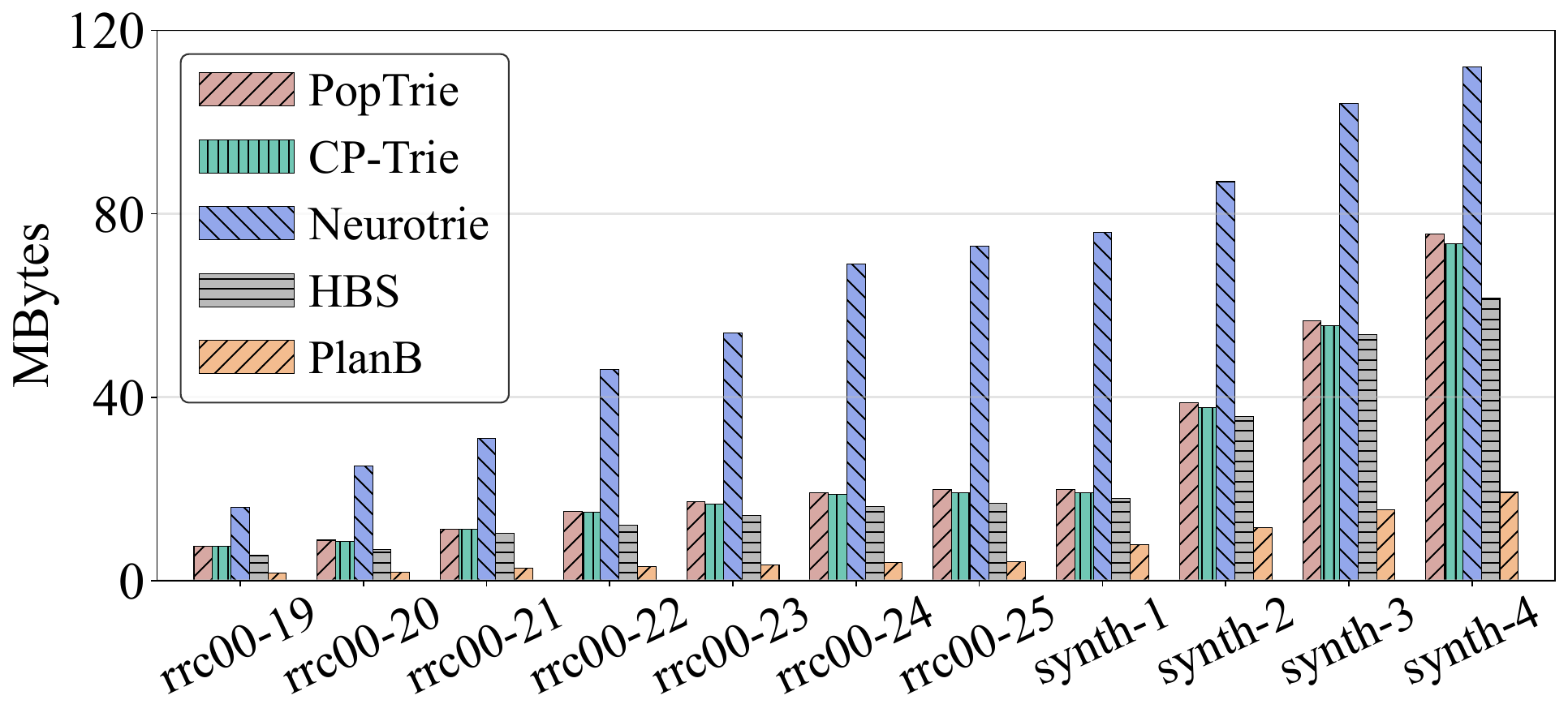}
    \vspace{-10pt}
    \caption{\label{fig:eva_mem} Memory overhead.
    }
    \vspace{-5pt}
\end{figure}

\re{As established in our design principles, maintaining cache residency is critical for high-speed lookups.
Fig. \ref{fig:eva_mem} quantifies \sys's advantage in this regard,
shows the memory overhead for various schemes across different FIBs.}
\re{Our evaluation shows that \sys reduces the memory overhead by 60.8\%$\sim$79.6\% compared to PopTrie,
59.3\%$\sim$79.7\% compared to CP-Trie,
82.8\%$\sim$92.5\% compared to Neurotrie,
and 56.4\%$\sim$75.9\% compared to HBS.}
\re{This dramatic reduction allows \sys's data structure to comfortably fit within the L3 cache of modern CPUs,
even for FIBs containing up to one million prefixes,
which is a key factor in its high throughput.}

By transforming prefixes into a set of elementary intervals,
\sys ensures that the size of its primary data structure scales linearly with the number of unique prefix boundaries (at most $2N+1$ for $N$ prefixes).
\re{This contrasts sharply with the overheads of competing methods:
trie-based schemes suffer from prefix expansion,
particularly for long IPv6 prefixes,
while HBS requires substantial extra space to maintain low hash collision rates.
\sys handles significantly larger FIBs before its performance is degraded by the memory hierarchy. While other schemes would face a performance collapse due to DRAM latency,
\sys maintains cache residency, ensuring high throughput and superior scalability.}

\begin{figure}[t]
    \begin{minipage}[t]{0.472\textwidth}
        \vfill
        \includegraphics[width=\linewidth]{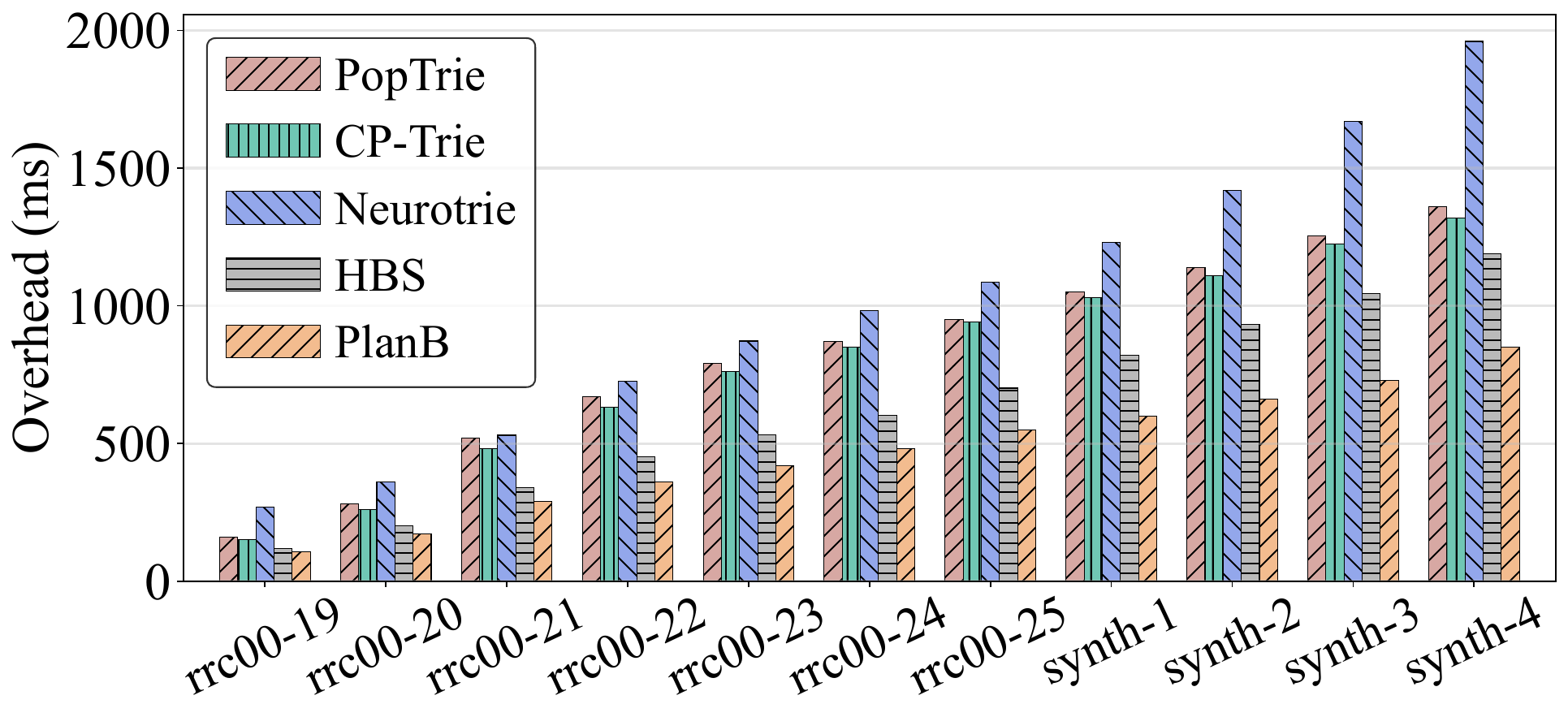}
        \vspace{-20pt}
        \caption{\label{fig:5_eva_update} Update overhead.
        }
    \end{minipage}
    \hfill
    \vspace{-5pt}
\end{figure}


\vspace{-3mm}
\subsection{Update Overhead}
\label{sec:update-overhead}

While prioritizing lookup performance,
\sys also maintains a highly efficient update mechanism.
By employing a batch-oriented, rebuild-and-swap strategy,
\sys achieves a lower amortized update overhead than competing schemes.
\re{As illustrated in Fig. \ref{fig:5_eva_update},
our evaluation across real-world FIBs shows that PlanB's update process is consistently faster,
reducing the average update overhead by 32.5\%$\sim$44.6\% compared to PopTrie,
28\%$\sim$41.5\% compared to CP-Trie,
49.3\%$\sim$60\% compared to Neurotrie,
and 10.2\%$\sim$21.5\% compared to HBS.
Neurotrie incurs high update overhead because rebuilding its structure requires executing a DRL-based inference process.
Moreover, for a FIB with 1 million prefixes (synth-4),
PlanB's total rebuild time is just 850 ms.
This quantification of the worst-case reconstruction overhead for large-scale datasets confirms that our rebuild-and-swap model remains practical and efficient even as routing tables continue to grow.}


\vspace{-3mm}
\subsection{Ablation Study of Design Components}
\label{sec:ablation-study}

\begin{figure}[t]
    \center
    \includegraphics[width=0.99\linewidth]{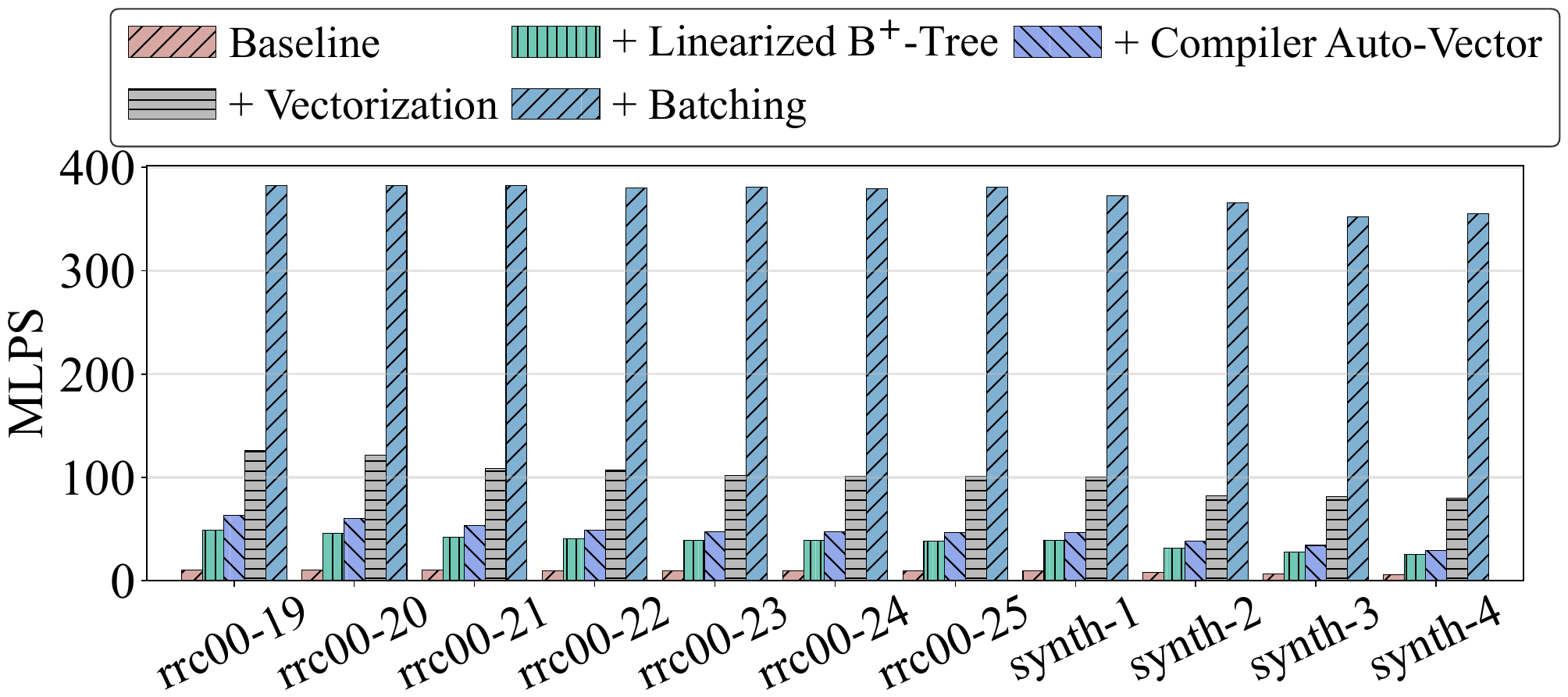}
    \vspace{-8pt}
    \caption{\label{fig:6_eva_planb} Ablation study of \sys's design components.
    }
    \vspace{-1pt}
\end{figure}

To quantify the performance contributions of \sys's design,
we conduct an ablation study on our AMD platform,
measuring single-core lookup speed as we progressively enable optimizations.
Fig. \ref{fig:6_eva_planb} illustrates the results across.

Our \textbf{Baseline}, a naive implementation using 1D interval transformation with standard C++ binary search (\texttt{std::lower\_bound}) on a sorted array,
achieves only 5.8$\sim$10.7 MLPS due to poor cache locality and branch mispredictions.
Replacing this with the \textbf{Linearized $B^+$-Tree and Branch-Free scalar logic} yields a speedup of 3.8$\times$$\sim$4.6$\times$ over the baseline.
We evaluate these two components together,
as compiler optimizations (-O3) automatically generate branch free code for a simple scalar loop.
Moreover, the sequential layout of the linearized $B^+$-tree inherently enables the subsequent zero-overhead vectorization and batching optimizations, underscoring its foundational role in the system.
\re{Enabling \textbf{Compiler Auto-Vectorization} for the in-node search provides another 1.15$\times$$\sim$1.3$\times$ speedup. This gain demonstrates the limitations of automatic compiler optimizations and serves as a reference for our manual approach.}


\re{Introducing AVX-512 \textbf{Vectorization} using intrinsics provides the next major boost,
outperforming the auto-vectorized version by 2$\times$$\sim$2.7$\times$.
This improvement comes from replacing the sequential in-node search with a single, parallel SIMD comparison.
Finally, incorporating \textbf{Batching} achieves \sys's full performance,
delivering an additional speedup of 3$\times$$\sim$4.5$\times$.
Batching effectively hides SIMD instruction latency by pipelining independent lookups, thus maximizing CPU utilization.
This study clearly demonstrates that each component of \sys's design contributes synergistically to its state-of-the-art performance.}

\vspace{-3mm}
\section{Discussion and Related Work}
\vspace{-1mm}

\textbf{Implementation on Diverse Hardware.}
\sys adapts easily to varied platforms.
Without 512-bit SIMD, wide comparisons simply decompose into narrower SIMD or scalar operations to build the bitmask. 
For hardware implementations on FPGAs or ASICs,
the linearized $B^+$-tree can reside in on-chip addressable SRAM.
The necessary operations,
simple aligned loads and vector to scalar comparisons,
are efficient and do not require a dedicated vector unit.
Moreover,
a hybrid software and hardware implementation can isolate the data plane from update overhead. A control plane processor can rebuild the lookup structure in software, while the data plane uses a hardware accelerator with the resulting linearized $B^+$-tree stored in on-chip memory for high speed forwarding.


\textbf{Near Constant Performance.}
\sys is highly resilient to unpredictable routing patterns and malicious attacks.
Unlike traditional traversal mechanisms that suffer severe performance degradation or potential router failure under random traffic due to overwhelmed pipelines and frequent cache misses,
\sys guarantees a fixed, shallow search depth.
By sustaining high cache-hit rates and branch-free execution,
\sys reliably maintains near constant throughput even under extreme dynamic workloads and security threats.

\textbf{Multiway Range Trees.}
Multiway Range Trees (MRTs) \cite{Multiway-Range-Trees-01,Multiway-Range-Trees-04} are a foundational IP lookup data structure that partitions the address space into elementary intervals.
MRTs are typically implemented as pointer-based B-trees.
Traversing these trees involves pointer chasing, which leads to memory indirection and can cause cache misses and pipeline stalls.
Moreover, MRT lookups are complex, requiring resolution of candidate prefixes stored in nodes along the root-to-leaf path.

\textbf{Hybrid CAM and RAM Architectures.}
Hybrid architectures leverage Content Addressable Memory (CAM) for fast, parallel lookups alongside high-density RAM. For instance, CRAM \cite{CRAM-NSDI} demonstrates that partitioning data structures across TCAM and SRAM supports databases exceeding individual memory capacities. Similarly, PtCAM \cite{PtCAM-SIGCOMM} accelerates name-prefix matching by storing compact Patricia trie bit-patterns in TCAM for the initial search, while offloading the final prefix verification to RAM.

\vspace{-3mm}
\section{Conclusion}
\vspace{-2mm}

We present \sys, a high-performance IPv6 lookup framework that recasts the LPM problem as a one-dimensional search problem over elementary intervals.
\sys traverses this simplified search space using linearized $B^+$-tree and a search algorithm aggressively optimized with SIMD instructions, batching, branch-free logic and loop unrolling.
These techniques fully exploit CPU parallelism to deliver high throughput.
Evaluation results show that \sys substantially outperforms current state-of-the-art software lookup solutions across a wide range of FIBs and hardware,
from mobile CPUs to server-grade processors.
%
\sys has been open-sourced
at \url{https://github.com/nicexlab/planb}.


\vspace{-4mm}
\section*{Acknowledgments}
\vspace{-2mm}
We thank Prof. Tao Li for his insightful discussion and key material support during the early stages of \sys.
We thank our shepherd, Prof. Jingxian Wang, and the anonymous reviewers for their valuable comments and suggestions.
Zhihao Zhang and Yiming Zhang are the co-primary authors.
This work is supported by the National Natural Science Foundation of China (grant no. 62441220).

\bibliographystyle{plain}
\bibliography{refs}

@ARTICLE{SAIL-ToN,
  author={Yang, Tong and Xie, Gaogang and Liu, Alex X. and Fu, Qiaobin and Li, Yanbiao and Li, Xiaoming and Mathy, Laurent},
  journal={IEEE/ACM Transactions on Networking},
  title={Constant IP Lookup With FIB Explosion}, 
  year={2018},
  pages={1821--1836}
  }

@inproceedings{SAIL-SIGCOMM,
author = {Yang, Tong and Xie, Gaogang and Li, YanBiao and Fu, Qiaobin and Liu, Alex X. and Li, Qi and Mathy, Laurent},
title = {Guarantee IP lookup performance with FIB explosion},
year = {2014},
booktitle = {Proceedings of ACM Conference on SIGCOMM (SIGCOMM'14)},
pages = {39--50},
}

@inproceedings{Poptrie,
author = {Asai, Hirochika and Ohara, Yasuhiro},
title = {Poptrie: A Compressed Trie with Population Count for Fast and Scalable Software IP Routing Table Lookup},
year = {2015},
booktitle = {Proceedings of the 2015 ACM Conference on Special Interest Group on Data Communication (SIGCOMM'15)},
pages = {57-70},
}

@inproceedings{HBS-APNet,
author = {Jiang, Donghong and Li, Yanbiao and Chen, Yuxuan and Hu, Jing and Huang, Yi and Xie, Gaogang},
title = {Heuristic Binary Search: Adaptive and Fast IPv6 Route Lookup with Incremental Updates},
year = {2023},
booktitle = {Proceedings of the 7th Asia-Pacific Workshop on Networking (APNet'23)},
pages = {47--53}
}

@ARTICLE{HBS-ToN,
  author={Jiang, Donghong and Li, Yanbiao and Chen, Yuxuan and Hu, Jing and Huang, Yi and Xie, Gaogang},
  journal={IEEE Transactions on Networking},
  title={Heuristic Binary Search: Adaptive and Fast IPv6 Route Lookup With Incremental Prefix Updates}, 
  year={2025},
  pages={554--569},
}

@ARTICLE{Neurotrie-ToN,
  author={Zhu, Yuxi and Chen, Hao and Yang, Yuan and Xu, Mingwei and Zhang, Yuxuan and Liu, Chenyi and Wu, Jianping},
  journal={IEEE/ACM Transactions on Networking}, 
  title={Fast Software IPv6 Lookup With Neurotrie}, 
  year={2024},
  pages={4040--4055},
}

@INPROCEEDINGS{Hi-BST,
  author={Shen, Tong and Yu, Xian and Xie, Gaogang and Zhang, Dafang},
  booktitle={2018 IEEE Global Communications Conference (GLOBECOM'18)},
  title={High-Performance IPv6 Lookup with Real-Time Updates Using Hierarchical-Balanced Search Tree}, 
  year={2018},
  pages={1--7},
}

@INPROCEEDINGS{CP-Trie,
  author={Islam, Md Iftakharul and Khan, Javed I},
  booktitle={2021 IEEE 22nd International Conference on High Performance Switching and Routing (HPSR'21)},
  title={CP-Trie: Cumulative PopCount based Trie for IPv6 Routing Table Lookup in Software and ASIC},
  year={2021},
  pages={1-8},
}

@INPROCEEDINGS{Li-ICICSP,
  author={Li, Yuqiang and Zhou, Fan and Zhu, Xiaoxiang and Liao, Jianming},
  booktitle={2022 5th International Conference on Information Communication and Signal Processing (ICICSP)}, 
  title={An IPv6 Routing Lookup Algorithm Based on Hash Table and HOT}, 
  year={2022},
  pages={397--402},
}

@inproceedings{SWAN,
author = {Hong, Chi-Yao and Kandula, Srikanth and Mahajan, Ratul and Zhang, Ming and Gill, Vijay and Nanduri, Mohan and Wattenhofer, Roger},
title = {Achieving high utilization with software-driven WAN},
year = {2013},
booktitle = {Proceedings of the ACM SIGCOMM 2013 Conference on SIGCOMM},
pages = {15--26},
}

@inproceedings {Orion-NSDI,
author = {Andrew D. Ferguson and Steve Gribble and Chi-Yao Hong and Charles Killian and Waqar Mohsin and Henrik Muehe and Joon Ong and Leon Poutievski and Arjun Singh and Lorenzo Vicisano and Richard Alimi and Shawn Shuoshuo Chen and Mike Conley and Subhasree Mandal and Karthik Nagaraj and Kondapa Naidu Bollineni and Amr Sabaa and Shidong Zhang and Min Zhu and Amin Vahdat},
title = {Orion: Google{\textquoteright}s {Software-Defined} Networking Control Plane},
booktitle = {18th USENIX Symposium on Networked Systems Design and Implementation (NSDI'21)},
year = {2021},
pages = {83--98}
}

@inproceedings{Kablan-HotMiddlebox,
author = {Kablan, Murad and Caldwell, Blake and Han, Richard and Jamjoom, Hani and Keller, Eric},
title = {Stateless Network Functions},
year = {2015},
booktitle = {Proceedings of the 2015 ACM SIGCOMM Workshop on Hot Topics in Middleboxes and Network Function Virtualization},
pages = {49--54},
}

@inproceedings{NFP-SIGCOMM,
author = {Sun, Chen and Bi, Jun and Zheng, Zhilong and Yu, Heng and Hu, Hongxin},
title = {NFP: Enabling Network Function Parallelism in NFV},
year = {2017},
booktitle = {Proceedings of the Conference of the ACM Special Interest Group on Data Communication (SIGCOMM'17)},
pages = {43--56},
}

@inproceedings{Bento-SIGCOMM,
author = {Reininger, Michael and Arora, Arushi and Herwig, Stephen and Francino, Nicholas and Hurst, Jayson and Garman, Christina and Levin, Dave},
title = {Bento: safely bringing network function virtualization to Tor},
year = {2021},
booktitle = {Proceedings of the 2021 ACM SIGCOMM 2021 Conference},
pages = {821--835},
}

@inproceedings{GreenNFV-SC,
author = {Nine, Md S. Q. Zulkar and Kosar, Tevfik and Bulut, Muhammed Fatih and Hwang, Jinho},
title = {GreenNFV: Energy-Efficient Network Function Virtualization with Service Level Agreement Constraints},
year = {2023},
booktitle = {Proceedings of the International Conference for High Performance Computing, Networking, Storage and Analysis (SC'23)},
}

@inproceedings {ResQ-NSDI,
author = {Amin Tootoonchian and Aurojit Panda and Chang Lan and Melvin Walls and Katerina Argyraki and Sylvia Ratnasamy and Scott Shenker},
title = {{ResQ}: Enabling {SLOs} in Network Function Virtualization},
booktitle = {15th USENIX Symposium on Networked Systems Design and Implementation (NSDI'18)},
year = {2018},
pages = {283--297},
}

@inproceedings{NeuroLPM-MICRO,
author = {Rashelbach, Alon and de Paula, Igor and Silberstein, Mark},
title = {NeuroLPM - Scaling Longest Prefix Match Hardware with Neural Networks},
year = {2023},
booktitle = {Proceedings of the 56th Annual IEEE/ACM International Symposium on Microarchitecture (MICRO'23)},
pages = {886--899},
}

@ARTICLE{TCache-ToN,
  author={Wan, Ying and Song, Haoyu and Xu, Yang and Wang, Yilun and Pan, Tian and Zhang, Chuwen and Wang, Yi and Liu, Bin},
  journal={IEEE/ACM Transactions on Networking}, 
  title={T-Cache: Efficient Policy-Based Forwarding Using Small TCAM}, 
  year={2021},
  pages={2693--2708},
  }

@inproceedings{Sadeh-SOSR,
author = {Sadeh, Yaniv and Rottenstreich, Ori and Kaplan, Haim},
title = {How Much TCAM do we Need for Splitting Traffic?},
year = {2021},
booktitle = {Proceedings of the ACM SIGCOMM Symposium on SDN Research (SOSR)},
pages = {169--175},
}

@inproceedings{TAR-APNet,
author = {Zhang, Xinyi and Xu, Zhiyuan and Zhao, Huaiyi and Li, Yanbiao and Xie, Gaogang},
title = {TAR: Traffic Adaptive IPv6 Routing Lookup Scheme},
year = {2024},
booktitle = {Proceedings of the 8th Asia-Pacific Workshop on Networking (APNet'24)},
pages = {135--141},
}

@article{Piraux-SIGCOMM,
author = {Piraux, Maxime and Barbette, Tom and Rybowski, Nicolas and Navarre, Louis and Alfroy, Thomas and Pelsser, Cristel and Michel, Fran\c{c}ois and Bonaventure, Olivier},
title = {The multiple roles that IPv6 addresses can play in today's internet},
year = {2022},
journal = {SIGCOMM Comput. Commun. Rev.},
pages = {10--18},
}

@inproceedings{Rye-SIGCOMM,
author = {Rye, Erik and Levin, Dave},
title = {IPv6 Hitlists at Scale: Be Careful What You Wish For},
year = {2023},
booktitle = {Proceedings of the ACM SIGCOMM 2023 Conference},
pages = {904--916},
}

@INPROCEEDINGS{Luori-ICNP,
  author={Cheng, Daguo and He, Lin and Wei, Chentian and Yin, Qilei and Jin, Boran and Wang, Zhaoan and Pan, Xiaoteng and Zhou, Sixu and Liu, Ying and Zhang, Shenglin and Tan, Fuchao and Liu, Wenmao},
  booktitle={2024 IEEE 32nd International Conference on Network Protocols (ICNP'24)},
  title={Luori: Active Probing and Evaluation of Internet-Wide IPv6 Fully Responsive Prefixes}, 
  year={2024},
  pages={1--12},
  }

@inproceedings {AddrMiner-ATC,
author = {Guanglei Song and Jiahai Yang and Lin He and Zhiliang Wang and Guo Li and Chenxin Duan and Yaozhong Liu and Zhongxiang Sun},
title = {{AddrMiner}: A Comprehensive Global Active {IPv6} Address Discovery System},
booktitle = {2022 USENIX Annual Technical Conference (ATC'22)},
year = {2022},
pages = {309--326},
}

@misc{RIPE-RIS,
  author       = {},
  title        = {{RIPE RIS is a BGP routing data collection platform}},
  howpublished = {\url{https://ris.ripe.net/docs/route-collectors/}}
}

@misc{RouteViews,
  author       = {},
  title        = {{University of Oregon RouteViews Project}},
  howpublished = {\url{https://www.routeviews.org/routeviews/collectors/}}
}

@misc{CNNIC-Report,
author = {China Internet Network Information Center (CNNIC)},
title = {The 55th Statistical Report on China's Internet Development},
howpublished = {\url{https://www.cnnic.com.cn/IDR/ReportDownloads/202505/P020250514564119130448.pdf}}
}

@misc{IPv6-Development-China,
author = {IPv6 Development in China},
title = {},
howpublished = {\url{https://www.cac.gov.cn/2025-08/1/c_1755590302116970.htm}}
}

@misc{IPv6-BGP-Data,
author = {IPv6 BGP Table Data},
title = {},
howpublished = {\url{https://bgp.potaroo.net/v6/as2.0/index.html}}
}

@inproceedings {Shao-NSDI,
author = {Hua Shao and Xiaoliang Wang and Yuanwei Lu and Yanbo Yu and Shengli Zheng and Youjian Zhao},
title = {Accessing Cloud with Disaggregated {Software-Defined} Router},
booktitle = {18th USENIX Symposium on Networked Systems Design and Implementation (NSDI'21)},
year = {2021},
pages = {1--14}
}

@inproceedings{RouteBricks-SOSP,
author = {Dobrescu, Mihai and Egi, Norbert and Argyraki, Katerina and Chun, Byung-Gon and Fall, Kevin and Iannaccone, Gianluca and Knies, Allan and Manesh, Maziar and Ratnasamy, Sylvia},
title = {RouteBricks: exploiting parallelism to scale software routers},
year = {2009},
booktitle = {Proceedings of the ACM SIGOPS 22nd Symposium on Operating Systems Principles (SOSP'09)},
pages = {15--28}
}

@inproceedings{casado2008rethinking,
  title={Rethinking Packet Forwarding Hardware.},
  author={Casado, Martin and Koponen, Teemu and Moon, Daekyeong and Shenker, Scott},
  booktitle={HotNets},
  pages={1--6},
  year={2008}
}

@INPROCEEDINGS{CATCAM-MICRO,
  author={Chen, Dibei and Li, Zhaoshi and Xiong, Tianzhu and Liu, Zhiwei and Yang, Jun and Yin, Shouyi and Wei, Shaojun and Liu, Leibo},
  booktitle={2020 53rd Annual IEEE/ACM International Symposium on Microarchitecture (MICRO'20)},
  title={CATCAM: Constant-time Alteration Ternary CAM with Scalable In-Memory Architecture}, 
  year={2020},
  pages={342--355}
}

@inproceedings{Mogul-PhysicalDeployability,
author = {Mogul, Jeffrey C. and Wilkes, John},
title = {Physical Deployability Matters},
year = {2023},
booktitle = {Proceedings of the 22nd ACM Workshop on Hot Topics in Networks (HotNets'23)},
pages = {9--17}
}

@inproceedings{Bhowmik-DEBS,
author = {Bhowmik, Sukanya and Tariq, Muhammad Adnan and Balogh, Alexander and Rothermel, Kurt},
title = {Addressing TCAM Limitations of Software-Defined Networks for Content-Based Routing},
year = {2017},
booktitle = {Proceedings of the 11th ACM International Conference on Distributed and Event-Based Systems (DEBS'17)},
pages = {100--111}
}

@INPROCEEDINGS{Jiang-INFOCOM,
  author={Jiang, W. and Wang, Q. and Prasanna, V. K.},
  booktitle={Proceedings of the IEEE INFOCOM 2008},
  title={Beyond TCAMs: An SRAM-Based Parallel Multi-Pipeline Architecture for Terabit IP Lookup}, 
  year={2008},
  pages={1786--1794},
}

@ARTICLE{Zheng-TON,
  author={Kai Zheng and Chengchen Hu and Hongbin Lu and Bin Liu},
  journal={IEEE/ACM Transactions on Networking}, 
  title={A TCAM-based distributed parallel IP lookup scheme and performance analysis}, 
  year={2006},
  volume={14},
  number={4},
  pages={863--875}
  }

@ARTICLE{Bando-ToN,
  author={Bando, Masanori and Lin, Yi-Li and Chao, H. Jonathan},
  journal={IEEE/ACM Transactions on Networking}, 
  title={FlashTrie: Beyond 100-Gb/s IP Route Lookup Using Hash-Based Prefix-Compressed Trie}, 
  year={2012},
  volume={20},
  number={4},
  pages={1262--1275}
  }

@INPROCEEDINGS{Stimpfling-CCGRID,
  author={Stimpfling, Thibaut and Langlois, J.M. Pierre and Bélanger, Normand and Savaria, Yvon},
  booktitle={2018 18th IEEE/ACM International Symposium on Cluster, Cloud and Grid Computing (CCGRID'18)}, 
  title={A Low-Latency Memory-Efficient IPv6 Lookup Engine Implemented on FPGA Using High-Level Synthesis}, 
  year={2018},
  pages={402--411}
}

@misc{Google-IPv6-Statistics,
author = {Google IPv6 Statistics},
title = {},
howpublished = {\url{https://www.google.com/intl/en/ipv6/statistics.html}}
}

@misc{APNIC-IPv6-Statistics,
author = {IPv6 Capable Rate by Country},
title = {},
howpublished = {\url{https://stats.labs.apnic.net/ipv6}}
}

@article{SHIP-ToN,
  title={SHIP: A scalable high-performance IPv6 lookup algorithm that exploits prefix characteristics},
  author={Stimpfling, Thibaut and Belanger, Normand and Langlois, JM Pierre and Savaria, Yvon},
  journal={IEEE/ACM Transactions on Networking},
  volume={27},
  number={4},
  pages={1529--1542},
  year={2019},
  publisher={IEEE}
}

@INPROCEEDINGS{DHL-ICDCS,
  author={Li, Xiaoyao and Wang, Xiuxiu and Liu, Fangming and Xu, Hong},
  booktitle={2018 IEEE 38th International Conference on Distributed Computing Systems (ICDCS'18)},
  title={DHL: Enabling Flexible Software Network Functions with FPGA Acceleration}, 
  year={2018},
  pages={1--11},
}

@INPROCEEDINGS{C2RTL-HPSR,
  author={Islam, Md Iftakharul and Khan, Javed I},
  booktitle={2021 IEEE 22nd International Conference on High Performance Switching and Routing (HPSR'21)},
  title={C2RTL: A High-level Synthesis System for IP Lookup and Packet Classification}, 
  year={2021},
  pages={1--8},
}

@inproceedings{ART-ICDE,
author = {Leis, Viktor and Kemper, Alfons and Neumann, Thomas},
title = {The adaptive radix tree: ARTful indexing for main-memory databases},
year = {2013},
booktitle = {Proceedings of the 2013 IEEE International Conference on Data Engineering (ICDE'13)},
pages = {38--49},
numpages = {12},
}

@article{Khuong-ACM-JEA,
author = {Khuong, Paul-Virak and Morin, Pat},
title = {Array Layouts for Comparison-Based Searching},
year = {2017},
volume = {22},
journal = {ACM J. Exp. Algorithmics},
month = may,
articleno = {1.3},
numpages = {39},
}

@misc{Intel-AVX512,
author = {Intel Architecture Instruction Set Extensions and Future Features Programming Reference},
title = {},
howpublished = {\url{https://www.intel.com/content/www/us/en/content-details/671368/intel-architecture-instruction-set-extensions-programming-reference.html}}
}

@misc{ARM-Neon,
author = {ARM NEON Technology},
title = {},
howpublished = {\url{https://www.arm.com/technologies/neon}}
}

@misc{DPDK-PerformanceThread,
author = {DPDK Performance Thread Model},
title = {},
howpublished = {\url{https://doc.dpdk.org/guides-16.04/sample_app_ug/performance_thread.html}}
}

@ARTICLE{Metronome-ToN,
  author={Faltelli, Marco and Belocchi, Giacomo and Quaglia, Francesco and Pontarelli, Salvatore and Bianchi, Giuseppe},
  journal={IEEE/ACM Transactions on Networking}, 
  title={Metronome: Adaptive and Precise Intermittent Packet Retrieval in DPDK}, 
  year={2023},
  volume={31},
  number={3},
  pages={979--993}
}

@misc{Apple-Silicon,
author = {Apple Silicon CPU Optimization Guide},
title = {},
howpublished = {\url{https://developer.apple.com/documentation/apple-silicon/cpu-optimization-guide}}
}

@article{Comer-ACM-CS,
author = {Comer, Douglas},
title = {Ubiquitous B-Tree},
year = {1979},
volume = {11},
number = {2},
journal = {ACM Comput. Surv.},
month = jun,
pages = {121--137},
numpages = {17}
}

@misc{Intel-SSE4,
author = {Intel SSE4 Programming Reference},
title = {},
howpublished = {\url{https://www.intel.com/content/dam/develop/external/us/en/documents/d9156103-705230.pdf}}
}

@inproceedings{CRAM-NSDI,
author = {Chang, Robert and Dogga, Pradeep and Fingerhut, Andy and Rios, Victor and Varghese, George},
title = {Scaling IP lookup to large databases using the CRAM lens},
year = {2025},
isbn = {978-1-939133-46-5},
publisher = {USENIX Association},
address = {USA},
booktitle = {Proceedings of the 22nd USENIX Symposium on Networked Systems Design and Implementation},
articleno = {8},
numpages = {20},
location = {Philadelphia, PA, USA},
series = {NSDI '25}
}

@inproceedings{PtCAM-SIGCOMM,
author = {Song, Tian and Li, Tianlong and Yang, Yating},
title = {PtCAM: Scalable High-Speed Name Prefix Lookup using TCAM},
year = {2025},
booktitle = {Proceedings of the ACM SIGCOMM 2025 Conference},
pages = {707--719},
numpages = {13},
location = {S\~{a}o Francisco Convent, Coimbra, Portugal},
series = {SIGCOMM '25}
}

@INPROCEEDINGS{Multiway-Range-Trees-01,
  author={Subhash Suri and Varghese, G. and Warkhede, P.R.},
  booktitle={GLOBECOM'01. IEEE Global Telecommunications Conference (Cat. No.01CH37270)}, 
  title={Multiway range trees: scalable IP lookup with fast updates},
  year={2001},
  volume={3},
  pages={1610-1614},
  }

@article{Multiway-Range-Trees-04,
author = {Warkhede, Priyank and Suri, Subhash and Varghese, George},
title = {Multiway range trees: scalable IP lookup with fast updates},
year = {2004},
volume = {44},
number = {3},
journal = {Comput. Netw.},
month = feb,
pages = {289--303},
}

@inproceedings{Sherman,
author = {Wang, Qing and Lu, Youyou and Shu, Jiwu},
title = {Sherman: A Write-Optimized Distributed B+Tree Index on Disaggregated Memory},
year = {2022},
booktitle = {Proceedings of the 2022 International Conference on Management of Data (SIGMOD'22)}, 
pages = {1033--1048},
}

@inproceedings{eurosys25deft,
author = {Wang, Jing and Wang, Qing and Zhang, Yuhao and Shu, Jiwu},
title = {{Deft: A Scalable Tree Index for Disaggregated Memory}},
year = {2025},
booktitle = {Proceedings of the Twentieth European Conference on Computer Systems (EuroSys'25)},
pages = {886--901}
}

@article{uTree,
author = {Chen, Youmin and Lu, Youyou and Fang, Kedong and Wang, Qing and Shu, Jiwu},
title = {uTree: a persistent B+-tree with low tail latency},
year = {2020},
journal = {Proc. VLDB Endow.},
month = jul,
pages = {2634--2648},
numpages = {15}
}






\end{document}